\documentclass[longauth]{aa}  
\usepackage[varg]{txfonts}
\usepackage{euclid}
\usepackage{booktabs} 
\usepackage{graphicx}
\usepackage{natbib}
\usepackage{orcidlink} 
\usepackage{tipa}
\usepackage{tabularx}
\hypersetup{
    colorlinks=true,
    linkcolor=blue,
    filecolor=magenta,      
    urlcolor=blue,
    citecolor=blue
}
\DeclareSIUnit\angstrom{\text {Å}}

\newcommand{\feii}{\ion{Fe}{II}\xspace}

\newcommand{\ciii}{\ion{C}{III}]~$\lambda$1909\xspace}

\newcommand{\nv}{\ion{N}{V}~$\lambda$1240\xspace}

\newcommand{\mgii}{\ion{Mg}{II}~$\lambda$2800\xspace}
\newcommand{\mgiishort}{\ion{Mg}{II}\xspace}

\newcommand{\oii}{\ionF{O}{II}{~$\lambda$3727}\xspace}

\newcommand{\oiiidd}{\ionF{O}{III}{~$\lambda\lambda$4959,5007}\xspace}
\newcommand{\oiiishort}{\ionF{O}{III}{}\xspace}

\newcommand{\hbetashort}{H\,$\beta$\xspace}

\newcommand{\hgamma}{H\,$\gamma$\xspace}
\newcommand{\civ}{\ion{C}{IV}~$\lambda$1549\xspace}
\newcommand{\civshort}{\ion{C}{IV}\xspace}
\newcommand{\lyafull}{Ly\,$\alpha$~$\lambda$1216\xspace}
\newcommand{\lya}{Ly\,$\alpha$\xspace}

\newcommand{\oiiiaur}{\ionF{O}{III}{~$\lambda$4363}\xspace}
\newcommand{\orcid}[1]{}

\begin{document} 

    \titlerunning{\textit{Euclid}: First discovery of a quasar at $z>8$}
\authorrunning{J. Wolf et al.}

\title{\textit{Euclid}: Discovery of a quasar at $z=8.13$ hosting a 360-million-solar-mass black hole \thanks{This paper is published on behalf of the Euclid Consortium.}}
	   
\author{J.~Wolf\orcid{0000-0003-0643-7935}\thanks{\email{julien.wolf@eso.org}}\inst{\ref{aff1},\ref{aff2},\ref{aff3}}
\and S.~Belladitta\orcid{0000-0003-4747-4484}\inst{\ref{aff1},\ref{aff4}}
\and D.~Yang\orcid{0000-0002-6769-0910}\inst{\ref{aff5}}
\and F.~Guarneri\orcid{0000-0003-4740-9762}\inst{\ref{aff6},\ref{aff7}}
\and E.~Ba\~nados\orcid{0000-0002-2931-7824}\inst{\ref{aff1}}
\and J.~F.~Hennawi\orcid{0000-0002-7054-4332}\inst{\ref{aff8},\ref{aff5}}
\and J.-T.~Schindler\orcid{0000-0002-4544-8242}\inst{\ref{aff6}}
\and R.~E.~Hviding\orcid{0000-0002-4684-9005}\inst{\ref{aff1}}
\and K.~Jahnke\orcid{0000-0003-3804-2137}\inst{\ref{aff1}}
\and D.~J.~Mortlock\orcid{0000-0002-0041-3783}\inst{\ref{aff9},\ref{aff10}}
\and X.~Fan\orcid{0000-0003-3310-0131}\inst{\ref{aff11}}
\and D.~Stern\orcid{0000-0003-2686-9241}\inst{\ref{aff12}}
\and A.-C.~Eilers\orcid{0000-0003-2895-6218}\inst{\ref{aff13},\ref{aff14}}
\and F.~Wang\orcid{0000-0002-7633-431X}\inst{\ref{aff15}}
\and J.~Yang\orcid{0000-0001-5287-4242}\inst{\ref{aff15}}
\and S.~E.~I.~Bosman\orcid{0000-0001-8582-7012}\inst{\ref{aff16},\ref{aff1}}
\and C.~J.~Conselice\orcid{0000-0003-1949-7638}\inst{\ref{aff17}}
\and F.~B.~Davies\orcid{0000-0003-0821-3644}\inst{\ref{aff1}}
\and R.~Decarli\orcid{0000-0002-2662-8803}\inst{\ref{aff4}}
\and A.~C.~N.~Hughes\orcid{0000-0001-9294-3089}\inst{\ref{aff9}}
\and I.~G.~Momcheva\orcid{0000-0003-1665-2073}\inst{\ref{aff1}}
\and M.~Onoue\orcid{0000-0003-2984-6803}\inst{\ref{aff18},\ref{aff19}}
\and L.~N.~Martinez-Ramirez\orcid{0009-0003-5506-5469}\inst{\ref{aff6}}
\and R.~A.~A.~Bowler\orcid{0000-0003-3917-1678}\inst{\ref{aff17}}
\and M.~Castellano\orcid{0000-0001-9875-8263}\inst{\ref{aff20}}
\and P.~Dayal\orcid{0000-0001-8460-1564}\inst{\ref{aff21},\ref{aff22},\ref{aff23}}
\and Y.~Matsuoka\orcid{0000-0001-5063-0340}\inst{\ref{aff24}}
\and Y.~Fu\orcid{0000-0002-0759-0504}\inst{\ref{aff5},\ref{aff25}}
\and H.~J.~A.~Rottgering\orcid{0000-0001-8887-2257}\inst{\ref{aff5}}
\and M.~Salvato\orcid{0000-0001-7116-9303}\inst{\ref{aff26}}
\and J.~Calhau\orcid{0000-0003-1803-6899}\inst{\ref{aff27}}
\and K.~I.~Caputi\orcid{0000-0001-8183-1460}\inst{\ref{aff25}}
\and G.~Castignani\orcid{0000-0001-6831-0687}\inst{\ref{aff4}}
\and Y.~Copin\orcid{0000-0002-5317-7518}\inst{\ref{aff28}}
\and A.~de~Graaff\orcid{0000-0002-2380-9801}\inst{\ref{aff1}}
\and J.~M.~Diego\orcid{0000-0001-9065-3926}\inst{\ref{aff29}}
\and P.~R.~M.~Eisenhardt\inst{\ref{aff12}}
\and A.~Fontana\orcid{0000-0003-3820-2823}\inst{\ref{aff20}}
\and C.~M.~Gutierrez\orcid{0000-0001-7854-783X}\inst{\ref{aff30},\ref{aff31}}
\and Y.~Harikane\orcid{0000-0002-6047-430X}\inst{\ref{aff32}}
\and H.~Hoekstra\orcid{0000-0002-0641-3231}\inst{\ref{aff5}}
\and R.~Maiolino\orcid{0000-0002-4985-3819}\inst{\ref{aff33}}
\and M.~Schirmer\orcid{0000-0003-2568-9994}\inst{\ref{aff1}}
\and D.~Scott\orcid{0000-0002-6878-9840}\inst{\ref{aff34}}
\and T.~Sekine\orcid{0009-0007-7660-6120}\inst{\ref{aff6}}
\and F.~Shankar\orcid{0000-0001-8973-5051}\inst{\ref{aff35}}
\and S.~Serjeant\orcid{0000-0002-0517-7943}\inst{\ref{aff36}}
\and L.~Spinoglio\orcid{0000-0001-8840-1551}\inst{\ref{aff37}}
\and G.~Vietri\orcid{0000-0001-9155-8875}\inst{\ref{aff38}}
\and F.~Walter\orcid{0000-0003-4793-7880}\inst{\ref{aff1}}
\and G.~Zamorani\orcid{0000-0002-2318-301X}\inst{\ref{aff4}}
\and S.~Andreon\orcid{0000-0002-2041-8784}\inst{\ref{aff39}}
\and C.~Baccigalupi\orcid{0000-0002-8211-1630}\inst{\ref{aff40},\ref{aff7},\ref{aff41},\ref{aff42}}
\and M.~Baldi\orcid{0000-0003-4145-1943}\inst{\ref{aff43},\ref{aff4},\ref{aff44}}
\and P.~Battaglia\orcid{0000-0002-7337-5909}\inst{\ref{aff4}}
\and R.~Bender\orcid{0000-0001-7179-0626}\inst{\ref{aff26},\ref{aff45}}
\and A.~Biviano\orcid{0000-0002-0857-0732}\inst{\ref{aff7},\ref{aff40}}
\and M.~Bolzonella\orcid{0000-0003-3278-4607}\inst{\ref{aff4}}
\and E.~Branchini\orcid{0000-0002-0808-6908}\inst{\ref{aff46},\ref{aff47},\ref{aff39}}
\and M.~Brescia\orcid{0000-0001-9506-5680}\inst{\ref{aff48},\ref{aff27}}
\and J.~Brinchmann\orcid{0000-0003-4359-8797}\inst{\ref{aff49},\ref{aff50},\ref{aff2}}
\and S.~Camera\orcid{0000-0003-3399-3574}\inst{\ref{aff51},\ref{aff52},\ref{aff53}}
\and G.~Ca\~nas-Herrera\orcid{0000-0003-2796-2149}\inst{\ref{aff5}}
\and V.~Capobianco\orcid{0000-0002-3309-7692}\inst{\ref{aff53}}
\and C.~Carbone\orcid{0000-0003-0125-3563}\inst{\ref{aff38}}
\and J.~Carretero\orcid{0000-0002-3130-0204}\inst{\ref{aff54},\ref{aff55}}
\and F.~J.~Castander\orcid{0000-0001-7316-4573}\inst{\ref{aff56},\ref{aff57}}
\and S.~Cavuoti\orcid{0000-0002-3787-4196}\inst{\ref{aff27},\ref{aff58}}
\and K.~C.~Chambers\orcid{0000-0001-6965-7789}\inst{\ref{aff59}}
\and A.~Cimatti\inst{\ref{aff60}}
\and C.~Colodro-Conde\inst{\ref{aff30}}
\and G.~Congedo\orcid{0000-0003-2508-0046}\inst{\ref{aff61}}
\and L.~Conversi\orcid{0000-0002-6710-8476}\inst{\ref{aff62},\ref{aff63}}
\and F.~Courbin\orcid{0000-0003-0758-6510}\inst{\ref{aff64},\ref{aff65},\ref{aff66}}
\and H.~M.~Courtois\orcid{0000-0003-0509-1776}\inst{\ref{aff67}}
\and M.~Cropper\orcid{0000-0003-4571-9468}\inst{\ref{aff68}}
\and J.-G.~Cuby\orcid{0000-0002-8767-1442}\inst{\ref{aff69},\ref{aff70}}
\and H.~Degaudenzi\orcid{0000-0002-5887-6799}\inst{\ref{aff71}}
\and G.~De~Lucia\orcid{0000-0002-6220-9104}\inst{\ref{aff7}}
\and H.~Dole\orcid{0000-0002-9767-3839}\inst{\ref{aff72}}
\and F.~Dubath\orcid{0000-0002-6533-2810}\inst{\ref{aff71}}
\and X.~Dupac\inst{\ref{aff63}}
\and S.~Escoffier\orcid{0000-0002-2847-7498}\inst{\ref{aff73}}
\and M.~Fabricius\orcid{0000-0002-7025-6058}\inst{\ref{aff26},\ref{aff45}}
\and M.~Farina\orcid{0000-0002-3089-7846}\inst{\ref{aff37}}
\and R.~Farinelli\inst{\ref{aff4}}
\and S.~Ferriol\inst{\ref{aff28}}
\and P.~Fosalba\orcid{0000-0002-1510-5214}\inst{\ref{aff57},\ref{aff56}}
\and S.~Fotopoulou\orcid{0000-0002-9686-254X}\inst{\ref{aff74}}
\and N.~Fourmanoit\orcid{0009-0005-6816-6925}\inst{\ref{aff73}}
\and M.~Fumana\orcid{0000-0001-6787-5950}\inst{\ref{aff38}}
\and L.~Gabarra\orcid{0000-0002-8486-8856}\inst{\ref{aff75}}
\and S.~Galeotta\orcid{0000-0002-3748-5115}\inst{\ref{aff7}}
\and K.~George\orcid{0000-0002-1734-8455}\inst{\ref{aff76}}
\and B.~Gillis\orcid{0000-0002-4478-1270}\inst{\ref{aff61}}
\and C.~Giocoli\orcid{0000-0002-9590-7961}\inst{\ref{aff4},\ref{aff44}}
\and J.~Gracia-Carpio\orcid{0000-0003-4689-3134}\inst{\ref{aff26}}
\and A.~Grazian\orcid{0000-0002-5688-0663}\inst{\ref{aff77}}
\and F.~Grupp\inst{\ref{aff26},\ref{aff45}}
\and S.~V.~H.~Haugan\orcid{0000-0001-9648-7260}\inst{\ref{aff78}}
\and S.~Hemmati\orcid{0000-0003-2226-5395}\inst{\ref{aff79}}
\and W.~Holmes\orcid{0009-0007-8554-4646}\inst{\ref{aff12}}
\and I.~M.~Hook\orcid{0000-0002-2960-978X}\inst{\ref{aff80}}
\and F.~Hormuth\inst{\ref{aff81}}
\and A.~Hornstrup\orcid{0000-0002-3363-0936}\inst{\ref{aff82},\ref{aff83}}
\and M.~Huertas-Company\orcid{0000-0002-1416-8483}\inst{\ref{aff30},\ref{aff84},\ref{aff85}}
\and M.~Jhabvala\inst{\ref{aff86}}
\and B.~Joachimi\orcid{0000-0001-7494-1303}\inst{\ref{aff87}}
\and S.~Kermiche\orcid{0000-0002-0302-5735}\inst{\ref{aff73}}
\and A.~Kiessling\orcid{0000-0002-2590-1273}\inst{\ref{aff12}}
\and B.~Kubik\orcid{0009-0006-5823-4880}\inst{\ref{aff28}}
\and K.~Kuijken\orcid{0000-0002-3827-0175}\inst{\ref{aff5}}
\and M.~K\"ummel\orcid{0000-0003-2791-2117}\inst{\ref{aff45}}
\and M.~Kunz\orcid{0000-0002-3052-7394}\inst{\ref{aff88}}
\and H.~Kurki-Suonio\orcid{0000-0002-4618-3063}\inst{\ref{aff89},\ref{aff90}}
\and R.~Laureijs\inst{\ref{aff25}}
\and A.~M.~C.~Le~Brun\orcid{0000-0002-0936-4594}\inst{\ref{aff91}}
\and V.~Le~Brun\orcid{0000-0002-5027-1939}\inst{\ref{aff70}}
\and S.~Ligori\orcid{0000-0003-4172-4606}\inst{\ref{aff53}}
\and P.~B.~Lilje\orcid{0000-0003-4324-7794}\inst{\ref{aff78}}
\and I.~Lloro\orcid{0000-0001-5966-1434}\inst{\ref{aff92}}
\and M.~Magliocchetti\orcid{0000-0001-9158-4838}\inst{\ref{aff37}}
\and G.~Mainetti\orcid{0000-0003-2384-2377}\inst{\ref{aff93}}
\and D.~Maino\inst{\ref{aff94},\ref{aff38},\ref{aff95}}
\and E.~Maiorano\orcid{0000-0003-2593-4355}\inst{\ref{aff4}}
\and O.~Mansutti\orcid{0000-0001-5758-4658}\inst{\ref{aff7}}
\and O.~Marggraf\orcid{0000-0001-7242-3852}\inst{\ref{aff96}}
\and M.~Martinelli\orcid{0000-0002-6943-7732}\inst{\ref{aff20},\ref{aff97}}
\and N.~Martinet\orcid{0000-0003-2786-7790}\inst{\ref{aff70}}
\and F.~Marulli\orcid{0000-0002-8850-0303}\inst{\ref{aff98},\ref{aff4},\ref{aff44}}
\and R.~J.~Massey\orcid{0000-0002-6085-3780}\inst{\ref{aff99}}
\and D.~C.~Masters\orcid{0000-0001-5382-6138}\inst{\ref{aff100}}
\and N.~Mauri\orcid{0000-0001-8196-1548}\inst{\ref{aff60},\ref{aff44}}
\and C.~J.~R.~McPartland\orcid{0000-0003-0639-025X}\inst{\ref{aff83},\ref{aff101}}
\and E.~Medinaceli\orcid{0000-0002-4040-7783}\inst{\ref{aff4}}
\and S.~Mei\orcid{0000-0002-2849-559X}\inst{\ref{aff102},\ref{aff103}}
\and M.~Meneghetti\orcid{0000-0003-1225-7084}\inst{\ref{aff4},\ref{aff44}}
\and E.~Merlin\orcid{0000-0001-6870-8900}\inst{\ref{aff77}}
\and G.~Meylan\orcid{0000-0001-6503-0209}\inst{\ref{aff104}}
\and P.~Monaco\orcid{0000-0003-2083-7564}\inst{\ref{aff105},\ref{aff7},\ref{aff41},\ref{aff40}}
\and A.~Mora\orcid{0000-0002-1922-8529}\inst{\ref{aff106}}
\and M.~Moresco\orcid{0000-0002-7616-7136}\inst{\ref{aff98},\ref{aff4}}
\and C.~Moretti\orcid{0000-0003-3314-8936}\inst{\ref{aff7},\ref{aff40},\ref{aff41}}
\and L.~Moscardini\orcid{0000-0002-3473-6716}\inst{\ref{aff98},\ref{aff4},\ref{aff44}}
\and R.~Nakajima\orcid{0009-0009-1213-7040}\inst{\ref{aff96}}
\and C.~Neissner\orcid{0000-0001-8524-4968}\inst{\ref{aff107},\ref{aff55}}
\and R.~C.~Nichol\orcid{0000-0003-0939-6518}\inst{\ref{aff108}}
\and S.-M.~Niemi\orcid{0009-0005-0247-0086}\inst{\ref{aff109}}
\and C.~Padilla\orcid{0000-0001-7951-0166}\inst{\ref{aff107}}
\and S.~Paltani\orcid{0000-0002-8108-9179}\inst{\ref{aff71}}
\and F.~Pasian\orcid{0000-0002-4869-3227}\inst{\ref{aff7}}
\and W.~J.~Percival\orcid{0000-0002-0644-5727}\inst{\ref{aff110},\ref{aff111},\ref{aff112}}
\and V.~Pettorino\orcid{0000-0002-4203-9320}\inst{\ref{aff109}}
\and S.~Pires\orcid{0000-0002-0249-2104}\inst{\ref{aff113}}
\and G.~Polenta\orcid{0000-0003-4067-9196}\inst{\ref{aff114}}
\and M.~Poncet\inst{\ref{aff115}}
\and L.~A.~Popa\inst{\ref{aff116}}
\and F.~Raison\orcid{0000-0002-7819-6918}\inst{\ref{aff26}}
\and R.~Rebolo\orcid{0000-0003-3767-7085}\inst{\ref{aff30},\ref{aff117},\ref{aff31}}
\and A.~Renzi\orcid{0000-0001-9856-1970}\inst{\ref{aff118},\ref{aff119},\ref{aff4}}
\and J.~Rhodes\orcid{0000-0002-4485-8549}\inst{\ref{aff12}}
\and G.~Riccio\inst{\ref{aff27}}
\and I.~Risso\orcid{0000-0003-2525-7761}\inst{\ref{aff46},\ref{aff47},\ref{aff39}}
\and H.-W.~Rix\orcid{0000-0003-4996-9069}\inst{\ref{aff1}}
\and F.~Rizzo\orcid{0000-0002-9407-585X}\inst{\ref{aff7}}
\and E.~Romelli\orcid{0000-0003-3069-9222}\inst{\ref{aff7}}
\and M.~Roncarelli\orcid{0000-0001-9587-7822}\inst{\ref{aff4}}
\and R.~Saglia\orcid{0000-0003-0378-7032}\inst{\ref{aff45},\ref{aff26}}
\and Z.~Sakr\orcid{0000-0002-4823-3757}\inst{\ref{aff120},\ref{aff121},\ref{aff122}}
\and D.~Sapone\orcid{0000-0001-7089-4503}\inst{\ref{aff123}}
\and M.~Sauvage\orcid{0000-0002-0809-2574}\inst{\ref{aff113}}
\and J.~A.~Schewtschenko\orcid{0000-0002-4913-6393}\inst{\ref{aff61}}
\and P.~Schneider\orcid{0000-0001-8561-2679}\inst{\ref{aff96}}
\and T.~Schrabback\orcid{0000-0002-6987-7834}\inst{\ref{aff124}}
\and A.~Secroun\orcid{0000-0003-0505-3710}\inst{\ref{aff73}}
\and E.~Sihvola\orcid{0000-0003-1804-7715}\inst{\ref{aff125}}
\and P.~Simon\inst{\ref{aff96}}
\and C.~Sirignano\orcid{0000-0002-0995-7146}\inst{\ref{aff118},\ref{aff119}}
\and G.~Sirri\orcid{0000-0003-2626-2853}\inst{\ref{aff44}}
\and L.~Stanco\orcid{0000-0002-9706-5104}\inst{\ref{aff119}}
\and P.~Tallada-Cresp\'{i}\orcid{0000-0002-1336-8328}\inst{\ref{aff54},\ref{aff55}}
\and A.~N.~Taylor\inst{\ref{aff61}}
\and I.~Tereno\orcid{0000-0002-4537-6218}\inst{\ref{aff126},\ref{aff127}}
\and N.~Tessore\orcid{0000-0002-9696-7931}\inst{\ref{aff68}}
\and S.~Toft\orcid{0000-0003-3631-7176}\inst{\ref{aff128},\ref{aff101}}
\and R.~Toledo-Moreo\orcid{0000-0002-2997-4859}\inst{\ref{aff129},\ref{aff130}}
\and F.~Torradeflot\orcid{0000-0003-1160-1517}\inst{\ref{aff55},\ref{aff54}}
\and I.~Tutusaus\orcid{0000-0002-3199-0399}\inst{\ref{aff56},\ref{aff57},\ref{aff121}}
\and E.~A.~Valentijn\inst{\ref{aff25}}
\and L.~Valenziano\orcid{0000-0002-1170-0104}\inst{\ref{aff4},\ref{aff131}}
\and J.~Valiviita\orcid{0000-0001-6225-3693}\inst{\ref{aff89},\ref{aff90}}
\and T.~Vassallo\orcid{0000-0001-6512-6358}\inst{\ref{aff7},\ref{aff76}}
\and D.~Vibert\orcid{0009-0008-0607-631X}\inst{\ref{aff70}}
\and Y.~Wang\orcid{0000-0002-4749-2984}\inst{\ref{aff79}}
\and J.~Weller\orcid{0000-0002-8282-2010}\inst{\ref{aff45},\ref{aff26}}
\and F.~M.~Zerbi\orcid{0000-0002-9996-973X}\inst{\ref{aff39}}
\and E.~Zucca\orcid{0000-0002-5845-8132}\inst{\ref{aff4}}
\and J.~Garc\'ia-Bellido\orcid{0000-0002-9370-8360}\inst{\ref{aff120}}}
										   
%%%% please do not edit the affiliation list -- contact ECEB Bureau for changes
\institute{Max-Planck-Institut f\"ur Astronomie, K\"onigstuhl 17, 69117 Heidelberg, Germany\label{aff1}
\and
European Southern Observatory, Karl-Schwarzschild-Str.~2, 85748 Garching, Germany\label{aff2}
\and
Excellence Cluster ORIGINS, Boltzmannstrasse 2, 85748 Garching, Germany\label{aff3}
\and
INAF-Osservatorio di Astrofisica e Scienza dello Spazio di Bologna, Via Piero Gobetti 93/3, 40129 Bologna, Italy\label{aff4}
\and
Leiden Observatory, Leiden University, Einsteinweg 55, 2333 CC Leiden, The Netherlands\label{aff5}
\and
Hamburger Sternwarte, University of Hamburg, Gojenbergsweg 112, 21029 Hamburg, Germany\label{aff6}
\and
INAF-Osservatorio Astronomico di Trieste, Via G. B. Tiepolo 11, 34143 Trieste, Italy\label{aff7}
\and
Department of Physics, University of California, Santa Barbara, CA 93106, USA\label{aff8}
\and
Astrophysics Group, Blackett Laboratory, Imperial College London, London SW7 2AZ, UK\label{aff9}
\and
Department of Mathematics, Imperial College London, London SW7 2AZ, UK\label{aff10}
\and
Steward Observatory, University of Arizona, 933 N. Cherry Ave, Tucson, AZ 85750, USA\label{aff11}
\and
Jet Propulsion Laboratory, California Institute of Technology, 4800 Oak Grove Drive, Pasadena, CA, 91109, USA\label{aff12}
\and
Department of Physics, Massachusetts Institute of Technology, Cambridge, MA 02139, USA\label{aff13}
\and
MIT Kavli Institute for Astrophysics and Space Research, Massachusetts Institute of Technology, Cambridge, MA 02139, USA\label{aff14}
\and
Department of Astronomy, University of Michigan, 1085 S. University Ave., Ann Arbor, MI 48109, USA\label{aff15}
\and
Institut f\"ur Theoretische Physik, University of Heidelberg, Philosophenweg 16, 69120 Heidelberg, Germany\label{aff16}
\and
Jodrell Bank Centre for Astrophysics, Department of Physics and Astronomy, University of Manchester, Oxford Road, Manchester M13 9PL, UK\label{aff17}
\and
Waseda Institute for Advanced Study (WIAS), Waseda University, 1-21-1, Nishi-Waseda, Shinjuku, Tokyo 169-0051, Japan\label{aff18}
\and
Kavli Institute for the Physics and Mathematics of the Universe (WPI), University of Tokyo, Kashiwa, Chiba 277-8583, Japan\label{aff19}
\and
INAF-Osservatorio Astronomico di Roma, Via Frascati 33, 00078 Monteporzio Catone, Italy\label{aff20}
\and
Canadian Institute for Theoretical Astrophysics, Toronto, ON M5S 3H8, Canada\label{aff21}
\and
David A. Dunlap Department of Astronomy \& Astrophysics, University of Toronto, 50 St George Street, Toronto, Ontario M5S 3H4, Canada\label{aff22}
\and
Department of Physics, 60 St George St, University of Toronto, Toronto, ON M5S 3H8, Canada\label{aff23}
\and
Research Center for Space and Cosmic Evolution, Ehime University, 2-5 Bunkyo-cho, Matsuyama, Ehime 790-8577, Japan\label{aff24}
\and
Kapteyn Astronomical Institute, University of Groningen, PO Box 800, 9700 AV Groningen, The Netherlands\label{aff25}
\and
Max Planck Institute for Extraterrestrial Physics, Giessenbachstr. 1, 85748 Garching, Germany\label{aff26}
\and
INAF-Osservatorio Astronomico di Capodimonte, Via Moiariello 16, 80131 Napoli, Italy\label{aff27}
\and
Universit\'e Claude Bernard Lyon 1, CNRS/IN2P3, IP2I Lyon, UMR 5822, Villeurbanne, F-69100, France\label{aff28}
\and
Instituto de F\'isica de Cantabria, Edificio Juan Jord\'a, Avenida de los Castros, 39005 Santander, Spain\label{aff29}
\and
Instituto de Astrof\'{\i}sica de Canarias, E-38205 La Laguna, Tenerife, Spain\label{aff30}
\and
Universidad de La Laguna, Dpto. Astrof\'\i sica, E-38206 La Laguna, Tenerife, Spain\label{aff31}
\and
Institute for Cosmic Ray Research, The University of Tokyo, 5-1-5 Kashiwanoha, Kashiwa, Chiba 277-8582, Japan\label{aff32}
\and
Cavendish Laboratory, University of Cambridge, JJ Thomson Avenue, Cambridge, CB3 0HE, UK\label{aff33}
\and
Department of Physics and Astronomy, University of British Columbia, Vancouver, BC V6T 1Z1, Canada\label{aff34}
\and
School of Physics \& Astronomy, University of Southampton, Highfield Campus, Southampton SO17 1BJ, UK\label{aff35}
\and
School of Physical Sciences, The Open University, Milton Keynes, MK7 6AA, UK\label{aff36}
\and
INAF-Istituto di Astrofisica e Planetologia Spaziali, via del Fosso del Cavaliere, 100, 00100 Roma, Italy\label{aff37}
\and
INAF-IASF Milano, Via Alfonso Corti 12, 20133 Milano, Italy\label{aff38}
\and
INAF-Osservatorio Astronomico di Brera, Via Brera 28, 20122 Milano, Italy\label{aff39}
\and
IFPU, Institute for Fundamental Physics of the Universe, via Beirut 2, 34151 Trieste, Italy\label{aff40}
\and
INFN, Sezione di Trieste, Via Valerio 2, 34127 Trieste TS, Italy\label{aff41}
\and
SISSA, International School for Advanced Studies, Via Bonomea 265, 34136 Trieste TS, Italy\label{aff42}
\and
Dipartimento di Fisica e Astronomia, Universit\`a di Bologna, Via Gobetti 93/2, 40129 Bologna, Italy\label{aff43}
\and
INFN-Sezione di Bologna, Viale Berti Pichat 6/2, 40127 Bologna, Italy\label{aff44}
\and
Universit\"ats-Sternwarte M\"unchen, Fakult\"at f\"ur Physik, Ludwig-Maximilians-Universit\"at M\"unchen, Scheinerstr.~1, 81679 M\"unchen, Germany\label{aff45}
\and
Dipartimento di Fisica, Universit\`a di Genova, Via Dodecaneso 33, 16146, Genova, Italy\label{aff46}
\and
INFN-Sezione di Genova, Via Dodecaneso 33, 16146, Genova, Italy\label{aff47}
\and
Department of Physics "E. Pancini", University Federico II, Via Cinthia 6, 80126, Napoli, Italy\label{aff48}
\and
Instituto de Astrof\'isica e Ci\^encias do Espa\c{c}o, Universidade do Porto, CAUP, Rua das Estrelas, PT4150-762 Porto, Portugal\label{aff49}
\and
Faculdade de Ci\^encias da Universidade do Porto, Rua do Campo de Alegre, 4150-007 Porto, Portugal\label{aff50}
\and
Dipartimento di Fisica, Universit\`a degli Studi di Torino, Via P. Giuria 1, 10125 Torino, Italy\label{aff51}
\and
INFN-Sezione di Torino, Via P. Giuria 1, 10125 Torino, Italy\label{aff52}
\and
INAF-Osservatorio Astrofisico di Torino, Via Osservatorio 20, 10025 Pino Torinese (TO), Italy\label{aff53}
\and
Centro de Investigaciones Energ\'eticas, Medioambientales y Tecnol\'ogicas (CIEMAT), Avenida Complutense 40, 28040 Madrid, Spain\label{aff54}
\and
Port d'Informaci\'{o} Cient\'{i}fica, Campus UAB, C. Albareda s/n, 08193 Bellaterra (Barcelona), Spain\label{aff55}
\and
Institute of Space Sciences (ICE, CSIC), Campus UAB, Carrer de Can Magrans, s/n, 08193 Barcelona, Spain\label{aff56}
\and
Institut d'Estudis Espacials de Catalunya (IEEC),  Edifici RDIT, Campus UPC, 08860 Castelldefels, Barcelona, Spain\label{aff57}
\and
INFN -- Sezione di Napoli, Via Cinthia 6, 80126, Napoli, Italy\label{aff58}
\and
Institute for Astronomy, University of Hawaii, 2680 Woodlawn Drive, Honolulu, HI 96822, USA\label{aff59}
\and
Dipartimento di Fisica e Astronomia "Augusto Righi" - Alma Mater Studiorum Universit\`a di Bologna, Viale Berti Pichat 6/2, 40127 Bologna, Italy\label{aff60}
\and
Institute for Astronomy, University of Edinburgh, Royal Observatory, Blackford Hill, Edinburgh EH9 3HJ, UK\label{aff61}
\and
European Space Agency/ESRIN, Largo Galileo Galilei 1, 00044 Frascati, Roma, Italy\label{aff62}
\and
ESAC/ESA, Camino Bajo del Castillo, s/n., Urb. Villafranca del Castillo, 28692 Villanueva de la Ca\~nada, Madrid, Spain\label{aff63}
\and
Institut de Ci\`{e}ncies del Cosmos (ICCUB), Universitat de Barcelona (IEEC-UB), Mart\'{i} i Franqu\`{e}s 1, 08028 Barcelona, Spain\label{aff64}
\and
Instituci\'o Catalana de Recerca i Estudis Avan\c{c}ats (ICREA), Passeig de Llu\'{\i}s Companys 23, 08010 Barcelona, Spain\label{aff65}
\and
Institut de Ciencies de l'Espai (IEEC-CSIC), Campus UAB, Carrer de Can Magrans, s/n Cerdanyola del Vall\'es, 08193 Barcelona, Spain\label{aff66}
\and
UCB Lyon 1, CNRS/IN2P3, IUF, IP2I Lyon, 4 rue Enrico Fermi, 69622 Villeurbanne, France\label{aff67}
\and
Mullard Space Science Laboratory, University College London, Holmbury St Mary, Dorking, Surrey RH5 6NT, UK\label{aff68}
\and
Canada-France-Hawaii Telescope, 65-1238 Mamalahoa Hwy, Kamuela, HI 96743, USA\label{aff69}
\and
Aix-Marseille Universit\'e, CNRS, CNES, LAM, Marseille, France\label{aff70}
\and
Department of Astronomy, University of Geneva, ch. d'Ecogia 16, 1290 Versoix, Switzerland\label{aff71}
\and
Universit\'e Paris-Saclay, CNRS, Institut d'astrophysique spatiale, 91405, Orsay, France\label{aff72}
\and
Aix-Marseille Universit\'e, CNRS/IN2P3, CPPM, Marseille, France\label{aff73}
\and
School of Physics, HH Wills Physics Laboratory, University of Bristol, Tyndall Avenue, Bristol, BS8 1TL, UK\label{aff74}
\and
Department of Physics, University of Oxford, Keble Road, Oxford OX1 3RH, UK\label{aff75}
\and
University Observatory, LMU Faculty of Physics, Scheinerstr.~1, 81679 Munich, Germany\label{aff76}
\and
INAF-Osservatorio Astronomico di Padova, Via dell'Osservatorio 5, 35122 Padova, Italy\label{aff77}
\and
Institute of Theoretical Astrophysics, University of Oslo, P.O. Box 1029 Blindern, 0315 Oslo, Norway\label{aff78}
\and
Caltech/IPAC, 1200 E. California Blvd., Pasadena, CA 91125, USA\label{aff79}
\and
Department of Physics, Lancaster University, Lancaster, LA1 4YB, UK\label{aff80}
\and
Felix Hormuth Engineering, Goethestr. 17, 69181 Leimen, Germany\label{aff81}
\and
Technical University of Denmark, Elektrovej 327, 2800 Kgs. Lyngby, Denmark\label{aff82}
\and
Cosmic Dawn Center (DAWN), Denmark\label{aff83}
\and
Universit\'e PSL, Observatoire de Paris, Sorbonne Universit\'e, CNRS, LERMA, 75014, Paris, France\label{aff84}
\and
Universit\'e Paris-Cit\'e, 5 Rue Thomas Mann, 75013, Paris, France\label{aff85}
\and
NASA Goddard Space Flight Center, Greenbelt, MD 20771, USA\label{aff86}
\and
Department of Physics and Astronomy, University College London, Gower Street, London WC1E 6BT, UK\label{aff87}
\and
Universit\'e de Gen\`eve, D\'epartement de Physique Th\'eorique and Centre for Astroparticle Physics, 24 quai Ernest-Ansermet, CH-1211 Gen\`eve 4, Switzerland\label{aff88}
\and
Department of Physics, P.O. Box 64, University of Helsinki, 00014 Helsinki, Finland\label{aff89}
\and
Helsinki Institute of Physics, Gustaf H{\"a}llstr{\"o}min katu 2, University of Helsinki, 00014 Helsinki, Finland\label{aff90}
\and
Laboratoire d'etude de l'Univers et des phenomenes eXtremes, Observatoire de Paris, Universit\'e PSL, Sorbonne Universit\'e, CNRS, 92190 Meudon, France\label{aff91}
\and
SKAO, Jodrell Bank, Lower Withington, Macclesfield SK11 9FT, UK\label{aff92}
\and
Centre de Calcul de l'IN2P3/CNRS, 21 avenue Pierre de Coubertin 69627 Villeurbanne Cedex, France\label{aff93}
\and
Dipartimento di Fisica "Aldo Pontremoli", Universit\`a degli Studi di Milano, Via Celoria 16, 20133 Milano, Italy\label{aff94}
\and
INFN-Sezione di Milano, Via Celoria 16, 20133 Milano, Italy\label{aff95}
\and
Universit\"at Bonn, Argelander-Institut f\"ur Astronomie, Auf dem H\"ugel 71, 53121 Bonn, Germany\label{aff96}
\and
INFN-Sezione di Roma, Piazzale Aldo Moro, 2 - c/o Dipartimento di Fisica, Edificio G. Marconi, 00185 Roma, Italy\label{aff97}
\and
Dipartimento di Fisica e Astronomia "Augusto Righi" - Alma Mater Studiorum Universit\`a di Bologna, via Piero Gobetti 93/2, 40129 Bologna, Italy\label{aff98}
\and
Department of Physics, Institute for Computational Cosmology, Durham University, South Road, Durham, DH1 3LE, UK\label{aff99}
\and
Infrared Processing and Analysis Center, California Institute of Technology, Pasadena, CA 91125, USA\label{aff100}
\and
Niels Bohr Institute, University of Copenhagen, Jagtvej 128, 2200 Copenhagen, Denmark\label{aff101}
\and
Universit\'e Paris Cit\'e, CNRS, Astroparticule et Cosmologie, 75013 Paris, France\label{aff102}
\and
CNRS-UCB International Research Laboratory, Centre Pierre Bin\'etruy, IRL2007, CPB-IN2P3, Berkeley, USA\label{aff103}
\and
Institute of Physics, Laboratory of Astrophysics, Ecole Polytechnique F\'ed\'erale de Lausanne (EPFL), Observatoire de Sauverny, 1290 Versoix, Switzerland\label{aff104}
\and
Dipartimento di Fisica - Sezione di Astronomia, Universit\`a di Trieste, Via Tiepolo 11, 34131 Trieste, Italy\label{aff105}
\and
Telespazio UK S.L. for European Space Agency (ESA), Camino bajo del Castillo, s/n, Urbanizacion Villafranca del Castillo, Villanueva de la Ca\~nada, 28692 Madrid, Spain\label{aff106}
\and
Institut de F\'{i}sica d'Altes Energies (IFAE), The Barcelona Institute of Science and Technology, Campus UAB, 08193 Bellaterra (Barcelona), Spain\label{aff107}
\and
School of Mathematics and Physics, University of Surrey, Guildford, Surrey, GU2 7XH, UK\label{aff108}
\and
European Space Agency/ESTEC, Keplerlaan 1, 2201 AZ Noordwijk, The Netherlands\label{aff109}
\and
Waterloo Centre for Astrophysics, University of Waterloo, Waterloo, Ontario N2L 3G1, Canada\label{aff110}
\and
Department of Physics and Astronomy, University of Waterloo, Waterloo, Ontario N2L 3G1, Canada\label{aff111}
\and
Perimeter Institute for Theoretical Physics, Waterloo, Ontario N2L 2Y5, Canada\label{aff112}
\and
Universit\'e Paris-Saclay, Universit\'e Paris Cit\'e, CEA, CNRS, AIM, 91191, Gif-sur-Yvette, France\label{aff113}
\and
Space Science Data Center, Italian Space Agency, via del Politecnico snc, 00133 Roma, Italy\label{aff114}
\and
Centre National d'Etudes Spatiales -- Centre spatial de Toulouse, 18 avenue Edouard Belin, 31401 Toulouse Cedex 9, France\label{aff115}
\and
Institute of Space Science, Str. Atomistilor, nr. 409 M\u{a}gurele, Ilfov, 077125, Romania\label{aff116}
\and
Consejo Superior de Investigaciones Cientificas, Calle Serrano 117, 28006 Madrid, Spain\label{aff117}
\and
Dipartimento di Fisica e Astronomia "G. Galilei", Universit\`a di Padova, Via Marzolo 8, 35131 Padova, Italy\label{aff118}
\and
INFN-Padova, Via Marzolo 8, 35131 Padova, Italy\label{aff119}
\and
Instituto de F\'isica Te\'orica UAM-CSIC, Campus de Cantoblanco, 28049 Madrid, Spain\label{aff120}
\and
Institut de Recherche en Astrophysique et Plan\'etologie (IRAP), Universit\'e de Toulouse, CNRS, UPS, CNES, 14 Av. Edouard Belin, 31400 Toulouse, France\label{aff121}
\and
Universit\'e St Joseph; Faculty of Sciences, Beirut, Lebanon\label{aff122}
\and
Departamento de F\'isica, FCFM, Universidad de Chile, Blanco Encalada 2008, Santiago, Chile\label{aff123}
\and
Universit\"at Innsbruck, Institut f\"ur Astro- und Teilchenphysik, Technikerstr. 25/8, 6020 Innsbruck, Austria\label{aff124}
\and
Department of Physics and Helsinki Institute of Physics, Gustaf H\"allstr\"omin katu 2, University of Helsinki, 00014 Helsinki, Finland\label{aff125}
\and
Departamento de F\'isica, Faculdade de Ci\^encias, Universidade de Lisboa, Edif\'icio C8, Campo Grande, PT1749-016 Lisboa, Portugal\label{aff126}
\and
Instituto de Astrof\'isica e Ci\^encias do Espa\c{c}o, Faculdade de Ci\^encias, Universidade de Lisboa, Tapada da Ajuda, 1349-018 Lisboa, Portugal\label{aff127}
\and
Cosmic Dawn Center (DAWN)\label{aff128}
\and
Universidad Polit\'ecnica de Cartagena, Departamento de Electr\'onica y Tecnolog\'ia de Computadoras,  Plaza del Hospital 1, 30202 Cartagena, Spain\label{aff129}
\and
European University of Technology EUt+, European Union\label{aff130}
\and
INFN-Bologna, Via Irnerio 46, 40126 Bologna, Italy\label{aff131}}

\date{Received July 24, 2026; accepted October 8, 2026}

\abstract{

Quasars powered by rapid accretion onto supermassive black holes (SMBHs) have been uncovered deep into the first gigayear of cosmic history. Here we present the discovery of EUCL\,J150452.291+625224.39 (J1504+6252), selected from Euclid Wide Survey imaging and spectroscopically confirmed as the first quasar at $z>8$ through a campaign with the Keck Observatory, the \textit{Hubble} Space Telescope, the Large Binocular Telescope, and the \textit{James Webb} Space Telescope (JWST). J1504+6252 exhibits the defining spectral signatures of archetypal quasars observed at a lower redshift, including a luminous blue power-law continuum and broad permitted emission lines, indicating the presence of a classical broad-line region. Using medium-resolution JWST/NIRSpec G395M spectroscopy, we determined a systemic redshift of $z_{\rm sys}=8.129\pm0.005$, corresponding to only $630\,\mathrm{Myr}$ after the Big Bang and shifting the quasar redshift frontier by $\simeq39\,\mathrm{Myr}$, or approximately $6\%$ of the age of the Universe at that epoch.
From the \hbetashort\ and \mgiishort\ emission lines, we measured intrinsic broad-line-region full widths at half maximum  FWHM(\hbetashort,BLR)=$3160^{+270}_{-270}\,\mathrm{km\,s^{-1}}$ and FWHM(\mgiishort,BLR)$=3740^{+160}_{-170}\,\mathrm{km\,s^{-1}}$, respectively. From a widely used single-epoch \hbetashort-based calibration, we measured a reference black hole mass of $\logten(M_{\rm BH}/M_\odot)=8.56^{+0.43}_{-0.43}$. Applying a range of single-epoch black hole mass calibrations to both \mgiishort and \hbetashort yields a broader interval of $\logten(M_{\rm BH}/M_\odot)=8.33$--$8.89$, with corresponding Eddington ratios of $\lambda_{\rm Edd}=0.12$--$0.44$. Assembling a black hole of several hundred million solar masses within the first few hundred million years, despite its currently sub-Eddington accretion rate, requires either a massive seed or sustained, highly efficient early growth. The discovery of a mature, sub-Eddington quasar at $z>8$ suggests that substantially more luminous quasars hosting similarly massive black holes may remain to be discovered. Conversely, the previously known re-ionisation-era quasars may represent the rapidly accreting end of a broader population of already massive black holes spanning a wider range of accretion states.
}


   \maketitle
   \nolinenumbers

\section{Introduction}

Quasars are the most luminous persistent sources of radiation in the Universe \citep[with absolute UV magnitudes of $M_{\rm UV}\lesssim -23$, e.g.][]{Banados2026}. The nuclear regions of quasars harbour actively accreting supermassive black holes (SMBHs), reaching $10^8$--$10^{10}\,M_\odot$ \citep[][and references therein]{Fan2023}.
Over the past three decades, the quasar redshift frontier has steadily advanced deep into the epoch of re-ionisation: from the first quasars beyond $z\approx5$ \citep{fan1999,fan2000, stern2000b}, to $z\approx6$ \citep{fan2001, fan2003, willott2007}, to the first quasar beyond  $z\approx7$ \citep{mortlock2011}, and, in rapid succession, three more at
$z\approx7.5$ \citep{banados2018a,Yang2020,wang2021}. Most recently, \citet{yang2026} pushed the frontier to $z=7.77$, corresponding to only about 660\,Myr %$\sim 663$\,Myr 
after the Big Bang, with the discovery of EUCL\,J172902.75$+$641018.1.

Together, these discoveries have extended our observational reach into the early Universe, in some cases by tens of millions of years. They have also opened several complementary channels for probing early cosmic history.

    First, extending the quasar redshift frontier to earlier  times places increasingly stringent constraints on black hole seeding and accretion models. The discovery of massive SMBHs at a progressively higher distance \citep[e.g.][]{onoue2019,yang2021,farina2022,bosman2025} leaves ever less time for their growth through conventional assembly pathways \citep{inayoshi2020}.
    
    Beyond constraining SMBH growth, as luminous background sources, the most distant quasars provide some of the longest sightlines through the intervening intergalactic medium (IGM), enabling studies of re-ionisation-era IGM opacity and metal enrichment \citep[e.g.][]{Davies2018a,tie24,hennawi25,kist2026}. Absorption features in their spectra, including damping-wing signatures and metal absorption systems \citep[e.g.][]{rowlands2026}, constrain the ionisation state, enrichment history, and large-scale opacity fluctuations of the IGM during the final stages of re-ionisation.
    
    Quasars further mark rare, highly biased peaks in the underlying matter distribution and therefore trace the densest regions of the early large-scale structure \citep[e.g.][]{Eilers2024,wangf2026,schindlerjt2026,Huang2026}. Their environments provide a unique view of the connection between the first rapidly growing SMBHs, massive galaxy overdensities, and the formation of protocluster-scale structure during the epoch of re-ionisation \citep[e.g.][]{pizzati2024,kapahtia2026}.
    
    Finally, on galactic scales, high-redshift quasars probe the early stages of SMBH--host coevolution \citep{kormendy2013}. Observations in the sub-millimitre regime reveal compact, $\lesssim 2$\,kpc dust-emitting regions, host dynamical masses of about $10^{10}$--$10^{11}\,M_\odot$, and frequent companions \citep[e.g.][]{Decarli2017,Decarli2018,venemans2020,Neeleman2021,belladitta2026}, while imaging with the \textit{James Webb} Space Telescope (JWST) Near-Infrared Camera \citep[NIRCam,][]{Rigby2023,rieke2023}  now enables direct stellar-mass measurements \citep[e.g.][]{Ding2023,Ding2025,Yue2024,Stone2024}. These constraints suggest a mean $M_{\rm BH}$--$M_\star$ relation broadly consistent with later epochs but with about 1\,dex scatter \citep{silverman2025}, permitting highly overmassive SMBHs in modest-mass hosts \citep[e.g.][]{stone2023,Yue2024}.

Luminous quasars in the epoch of re-ionisation are exceptionally rare: at $z\simeq7$,
the cumulative co-moving space density is of the order of one source per Gpc$^3$ for quasars brighter than $M_{1450}\sim-25$, corresponding to a projected surface density of only one such quasar per about 130\,deg$^2$, and it declines rapidly towards higher luminosities \citep[e.g.][]{schindler2023,matsuoka2023}. To date, only seven quasars have been spectroscopically confirmed at $z>7.5$ \citep[][]{banados2018a,Yang2020,wang2021,yang2026}. All were identified as optical dropouts in wide-area optical and near-infrared (NIR) imaging surveys: at these redshifts, neutral hydrogen absorption in the intervening IGM strongly suppresses flux bluewards of \lyafull (hereafter Ly$\,\alpha$), producing a sharp optical-to-NIR colour break. 
 
The \Euclid\ mission \citep{EuclidSkyOverview} started its wide survey in February 2024, following the launch of the spacecraft in July 2023. 
Using the first 1.5\,yr of imaging survey data over approximately $3000\,{\rm deg}^2$, \citet{yang2026} spectroscopically confirmed 31 new quasars at $6.60<z<7.77$, increasing the number of known $z>7.5$ quasars from three to seven. These objects were spectroscopically confirmed with ground-based observations on the Keck and Magellan telescopes.

The high cost of deep ground-based NIR spectroscopy and telluric absorption limit the accessible redshift and magnitude range. For instance, the faintest quasar \citet{yang2026} were able to spectroscopically confirm is EUCL\,J101255.87+663058.0 with \JE$=23.0$ (see Sect. \ref{sec:ews}). Its confirmation required a 7800\,s exposure with the Keck/Multi-Object Spectrograph for Infra-Red Exploration \citep[MOSFIRE,][]{mclean2010,mclean2012}. 

For these reasons, we have decided to pursue space-based follow-up programmes for faint \Euclid-identified high-redshift quasar candidates. The \textit{Hubble} Space Telescope (HST) Wide Field Camera 3 \citep[WFC3; ][]{kimble2008} GO 18112 programme\footnote{\url{https://www.stsci.edu/hst-program-info/program/?program=18112}} (PI: J.~Wolf) and the JWST Near Infrared Spectrograph \citep[NIRSpec; ][] {Jakobsen2022} GO 11173 programme\footnote{\url{https://www.stsci.edu/jwst/science-execution/program-information?id=11173}} (PI: J.-T.~Schindler) were designed to overcome the limits of ground-based observations.

Here we report the discovery of EUCL\,J150452.291+625224.39 (hereafter J1504+6252, nicknamed Dún Aonghasa), the first quasar confirmed at $z>8$ and hence the most distant quasar currently known\footnote{We name this source Dún Aonghasa (Irish pronunciation: /\textipa{du:n "e:Ng@s@}/), after the prehistoric cliff-top fort on Inis Mór. Built at the Atlantic edge of Ireland, where land gives way to a vast oceanic horizon to the west, Dún Aonghasa evokes both a boundary and an invitation: an  apparent end at the edge of the cliff, with much still hidden beyond it. Likewise, J1504+6252 stands at the current observational frontier of quasars, illuminating the threshold to even earlier times still waiting to be explored.}.
It was identified in the Euclid Wide Survey (EWS) and we describe the selection process in Sect.~\ref{sec:sel}. In Sect.~\ref{sec:spec}, we present its confirmation through NIR spectroscopic observations with Keck/MOSFIRE, HST/WFC3, the Large Binocular Telescope (LBT) Utility Camera in the Infrared \citep[LUCI,][]{mandel2000,seifert2003}, and JWST/NIRSpec. In Sect.~\ref{sec:bhm}, we present the derivation of the nuclear properties from the JWST spectrum. In Sect.~\ref{sec:disc}, we discuss this discovery in the context of early SMBH growth and the emergence of the first massive black holes. Finally, we present our conclusions in  Sect.~\ref{sec:conclusions}. Throughout this paper, we adopt the Planck 2018 cosmology from \citet{Planck2020} for all distance-dependent measurements, with $H_0=67.66\,{\rm km\,s^{-1}\,Mpc^{-1}}$, $\Omega_{\rm m}=0.3111$, and $\Omega_\Lambda=0.6889$, and use magnitudes with an AB zero-point \citep{oke1983}. Unless stated otherwise, quoted values correspond to the posterior median, with uncertainties given by the $16$th and $84$th percentiles, representing equal-tailed $68\%$ credible intervals.

\section{Candidate selection}

\label{sec:sel}

J1504+6252 was identified as a high-probability high-$z$ quasar candidate in EWS imaging using a combination of advanced selection techniques and redshift estimate. Here we detail the identification and photometric redshift determination of the source from custom \Euclid catalogues. 

\subsection{Deep imaging from the EWS}
\label{sec:ews}
The EWS will cover about 14\,000\,deg$^2$ of the extragalactic sky over the nominal six-year mission \citep{EuclidSkyOverview}. Imaging is obtained with the Visible Camera \citep[VIS,][]{EuclidSkyVIS}, providing a single broad optical band, \IE, over $550$--$900\,{\rm nm}$, and with the Near Infrared Spectrometer and Photometer \citep[NISP,][]{EuclidSkyNISP}, which provides NIR imaging in three bands \citep[][]{Schirmer-EP18}: \YE ($949.6$--$1212.3\,{\rm nm}$), \JE ($1167.6$--$1567.0\,{\rm nm}$), and \HE ($1521.5$--$2021.4\,{\rm nm}$). The nominal EWS depths are $\IE\simeq26.2$ and \YE, \JE, \HE $\simeq$ 24--24.5, quoted as $5~\sigma$ point-source limits.

\subsection{Candidate selection}

We built a custom candidate catalogue from \Euclid VIS and NISP imaging. The detailed catalogue construction process is presented by \citet{yang2026}. Here, we only summarise the most important steps. First, we selected isolated, point-like NIR sources detected at signal-to-noise ratio (${\rm S/N}$) $>5$ in either \JE or \HE, and then applied an optical-dropout criterion using VIS photometry. Force point-spread-function (PSF) photometry was measured on either the native VIS/NISP stacks or, when available, the co-added MER mosaics, which assemble calibrated exposures into co-aligned, background-subtracted images on a common pixel grid \citep{Q1-TP004}. The final dropout sample was defined by the colour criterion $\IE-\JE>3.05$ using the newly measured PSF photometry before passing candidates to the quasar-selection pipelines.

To improve completeness and candidate reliability, the dropout catalogues were submitted to three independent quasar selection algorithms. The methodologies and method-specific success metrics are detailed in \citet{yang2026}; the three methods are:

\begin{itemize}
    \item Extreme Deconvolution High-Redshift Quasar selection \citep[XDHZQSO,][]{nanni2022,yang2024,yang2026} modelling the quasar and contaminant populations in flux-ratio space, accounting for photometric uncertainties and their dependence on magnitude. 
    \item  An XGBoost classifier \citep[][]{chenguestrin2016} trained on simulated high-redshift quasars \citep[][]{guarneri2016} and empirical contaminants to assign each source a quasar probability.
    \item  
    Template-based spectral energy distribution (SED) fitting with \texttt{LePhare} \citep[v2.2][]{lephare2011}, comparing quasar, galaxy, and stellar-contaminant solutions \citep[][]{wolf24}.
\end{itemize}
  Each method returned a numeric selection success metric and a photometric redshift estimate. The candidates that passed empirical thresholds were visually inspected to remove data artefacts or obvious astrophysical contaminants and were merged into a prioritised target list for spectroscopic follow-up.

\begin{figure}
    \centering
    \includegraphics[width=1.0\linewidth]{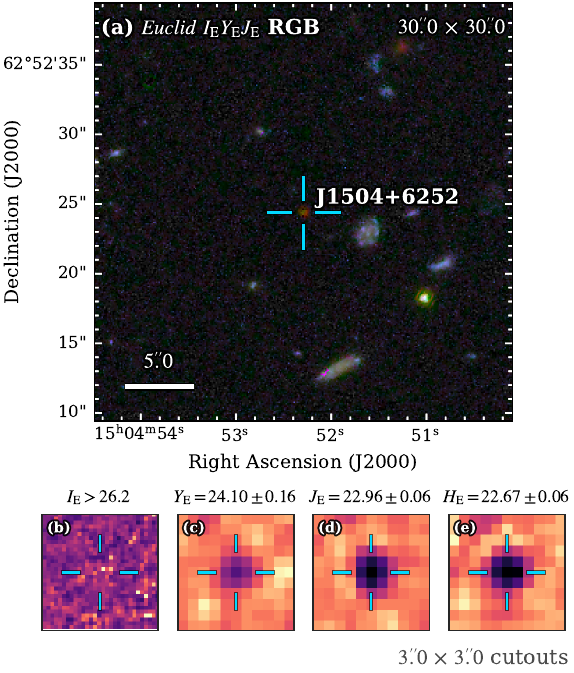}
    \caption{\Euclid imaging of J1504+6252. Panel (a) shows a $30^{\prime\prime}\times30^{\prime\prime}$ \Euclid RGB image using \IE-, \YE-, and \JE-band MER cutouts centred on the quasar, whose position is indicated by the cyan crosshairs. Panels (b)--(e) show $3^{\prime\prime}\times3^{\prime\prime}$ cutouts of the VIS and NISP stacks in the \IE, \YE, \JE, and \HE bands, respectively, with cyan crosshairs marking the source position. J1504+6252 is undetected in the broad optical \IE band, is suppressed in \YE, and is clearly detected in \JE and \HE, producing the sharp spectral break expected for a quasar at $z>8$.
}
    \label{fig:imaging}
\end{figure}

J1504+6252 was selected by all three methods as a high-priority candidate. 
It was identified within the 3644\,deg$^2$ observed region of the survey available in August 2025 
and has measured PSF photometry $\YE=24.10\pm0.16$, $\JE=22.96\pm0.06$, and $\HE=22.67\pm0.06$ (see Table~\ref{tab:j1504_properties}), with no significant detection in the optical \IE band down to the nominal (5~$\sigma$) point-source depth of \IE$\simeq26.2$ mag. The source is not covered by ancillary $z$-band imaging. The photometric redshift estimates from XDHZQSO, XGBoost, and SED fitting are respectively $z_{\rm phot}\in \{8.20, 8.21,7.98\}$. Combining the three estimates gives a mean photometric redshift of $\Bar{z}_{\rm phot}=8.13\pm0.10$. This value informed the grism assignment for spectroscopic follow-up (see Sect. \ref{sec:hst}). \Euclid cutouts of the source, together with a $30^{\prime\prime}$-wide view of its surrounding field\footnote{The composite image was produced with \texttt{eummy} \citep{Schirmer2026eummy}, which combines the \Euclid imaging channels to preserve the VIS spatial resolution while optimally encoding the multi-band colour information.}, are shown in Fig.~\ref{fig:imaging}. The object appears as an isolated red point source, with no evidence for multiple images, arcs, or an obvious deflector; the nearest foreground galaxy lies at a projected separation of $\sim5^{\prime\prime}$. Strong gravitational lensing is therefore unlikely, although modest, unresolved magnification cannot be excluded based on the available \Euclid imaging.

\section{Quasar confirmation}
\label{sec:spec}

The photometric redshift, $\bar{z}_{\rm phot}\simeq8.1$, and faintness of J1504+6252 motivated a tiered spectroscopic campaign combining deep ground- and space-based observations, each providing progressively stronger evidence for its quasar nature and redshift.
 An initial Keck/MOSFIRE long-slit spectrum revealed a blue continuum (Sect.~\ref{sec:keck}), while HST/WFC3 provided tentative evidence for a spectral break (Sect.~\ref{sec:hst}). This break and a flat continuum were subsequently established with a deep LBT/LUCI exposure (Sect.~\ref{sec:luci}), and a final JWST/NIRSpec observation secured the redshift and delivered a high-quality, broad-wavelength-coverage spectrum for further analysis (Sect.~\ref{sec:jwst}). The spectroscopic campaign is summarised in Table \ref{tab:campaign}.  We did not use the \Euclid/NISP slitless spectroscopic channel because, at \JE$\simeq23$, the source is well below the expected continuum-detection limit of the EWS \citep{EP-Gillard}.
 
\begin{table}
\caption{Summary of the spectroscopic observations of J1504+6252.}
\label{tab:observations}
\centering
\small
\setlength{\tabcolsep}{3.5pt}
\begin{tabularx}{\columnwidth}{@{}llXlc@{}}
\toprule
\toprule
Telescope &
Instrument &
Grating/grism and filter &
Date &
$t_{\rm exp}$ [s] \\
\midrule
Keck I &
MOSFIRE &
J2 &
2026-01-18 &
4800 \\

HST &
WFC3/IR &
G141/F125W &
2026-05-12 &
2206 \\

LBT &
LUCI1+LUCI2 &
G200/HKspec &
2026-05-15 &
21\,360 \\

LBT &
LUCI1+LUCI2 &
G200/zJspec &
2026-05-16 &
23\,520 \\

JWST &
NIRSpec &
G140H/F100LP &
2026-06-26 &
788 \\

JWST &
NIRSpec &
G235H/F170LP &
2026-06-26 &
591 \\

JWST &
NIRSpec &
G395M/F290LP &
2026-06-26 &
788 \\
\bottomrule
\end{tabularx}

\tablefoot{
The listed exposure times are total on-source integration times.
For the LBT observations, they refer to the retained exposures combined
from LUCI1 and LUCI2. The WFC3/F125W image was obtained for source
positioning and wavelength calibration of the G141 slitless spectrum.
}
\label{tab:campaign}
\end{table}

\subsection{Keck/MOSFIRE longslit spectroscopy}
\label{sec:keck}
Initial spectroscopic follow-up was obtained with Keck/MOSFIRE on 2026-01-18 (U262, PI: J.~Hennawi). The observations were carried out in longslit mode with the J2 filter, covering $\lambda_{\rm obs}\simeq1120$--$1250\,{\rm nm}$. We employed an ABAB dithering pattern for sky subtraction, with a total on-source exposure time of 4800\,s. The MOSFIRE data were reduced with \texttt{PypeIt} \citep[][version 1.17.3]{pypeit:zenodo,pypeit:joss_pub} following the same procedure described in \cite{yang2026}.

The MOSFIRE spectrum shows a blue continuum detection down to the blue edge of the J2 bandpass (50\% transmission at 1117\,nm), consistent with a high-$z$ quasar with the observed \JE magnitude. However, a spectral break was not covered within the observed range, constraining the redshift to $z\lesssim 8.2$. While these data could not, by themselves, localise the break, they established J1504+6252 as a promising $z\gtrsim8$ quasar candidate, consistent with the photometric redshift.

\subsection{HST/WFC3 slitless spectroscopy}
\label{sec:hst}
HST GO 18112 (PI: J.~Wolf) is a WFC3/IR slitless-spectroscopy snapshot programme designed to obtain spectra of \Euclid high-redshift quasar candidates within a single HST orbit. The programme uses the WFC3/IR grisms G102 and G141, which cover 800--1150\,nm and 1075--1700\,nm, respectively. For each target, the grism setup is selected from the candidate photometric redshift: candidates with $z_{\rm phot,mean}-s_{\rm phot}<8.0$ are observed with F105W pre-imaging and the G102 grism, while candidates with $z_{\rm phot,mean}-s_{\rm phot}>8.0$ are observed with F125W pre-imaging and the G141 grism, to capture the \lya-break.\footnote{Here $z_{\rm phot,mean}$ denotes the mean photometric redshift from the three selection methods. $s_{\rm phot}$ is the standard deviation and is a measure of the dispersion among the methods. This quantity is only used for grism selection; it is not a formal photometric-redshift uncertainty.}

J1504+6252 was observed as part of this programme on 2026-05-12, at the angle between the U3 axis of the telescope relative to north ORIENT(U3) $=177^\circ$, using the WFC3/IR G141 grism. The visit consisted of a direct F125W image with an exposure time of 113\,s, followed by two dithered G141 spectroscopic exposures of 1103\,s each, for a total grism exposure time of 2206\,s.

The slitless spectroscopic data were reduced with \texttt{Grizli} \citep[][see for instance \citealt{watson2025} for the detail of the grism reduction]{grizli}. We began with the standard calibrated HST exposures and applied additional preprocessing to improve the quality of the direct and grism images, including astrometric alignment, masking of bad pixels and cosmic rays, and construction of combined direct-image mosaics used for source detection. The grism exposures were corrected for the spatially structured NIR background using scaled master-background models, with residual large-scale structure removed after masking object flux. Object catalogues and segmentation maps derived from the direct F125W image were then used to predict the position, trace, and wavelength solution of each dispersed spectrum. \texttt{Grizli} was used to model the full two-dimensional grism scene, including contamination from neighbouring sources, and to subtract this contamination before extracting individual spectra. The final products consist of contamination-corrected two-dimensional spectra and optimally extracted one-dimensional spectra, which were used for subsequent redshift and emission-line measurements.

The resulting 2D and 1D spectra are presented in Fig.~\ref{fig:spec}. The HST/WFC3 G141 spectrum reveals a clear continuum detection redwards of a sharp break at $\lambda_{\rm obs}\simeq1120\,{\rm nm}$, as expected from a Ly$\,\alpha$ absorption break at $z\simeq8.1$. We fit the WFC3/G141 low-resolution spectrum with quasar templates following the approach of \citet{banados2023}, \citet{belladitta2025}, and \citet{martinez2026}, and obtain a best-fit redshift of $z=8.09$ with $\chi^2=45.6$ (53 degrees of freedom).
The G141 grism resolution, $R\simeq130$, corresponds to $\Delta\lambda\simeq8$--$10\,{\rm nm}$ at this wavelength, or $\Delta z\simeq0.07$ for Ly$\,\alpha$; allowing for the proximity of the break to the blue sensitivity limit and the uncertain break morphology, we quote a grism-based estimate of $z_{\rm grism}\simeq8.1\pm0.1$. In addition, the spectrum shows a broad emission feature near $\lambda_{\rm obs}\simeq1430\,{\rm nm}$, consistent with \civ (hereafter \civshort) at the same redshift. The combined presence of the Ly$\,\alpha$ break and broad high-ionisation emission identifies J1504+6252 as a quasar at $z\simeq8.1$, confirming the \Euclid-based photometric redshift. However, because the Ly$\,\alpha$ break lies close to the blue sensitivity limit of the G141 grism, additional spectroscopic follow-up was required to confirm the break and refine the redshift.

\subsection{LBT/LUCI longslit spectroscopy}
\label{sec:luci}
\begin{figure*}
    \centering
    \includegraphics[width=1.0\linewidth]{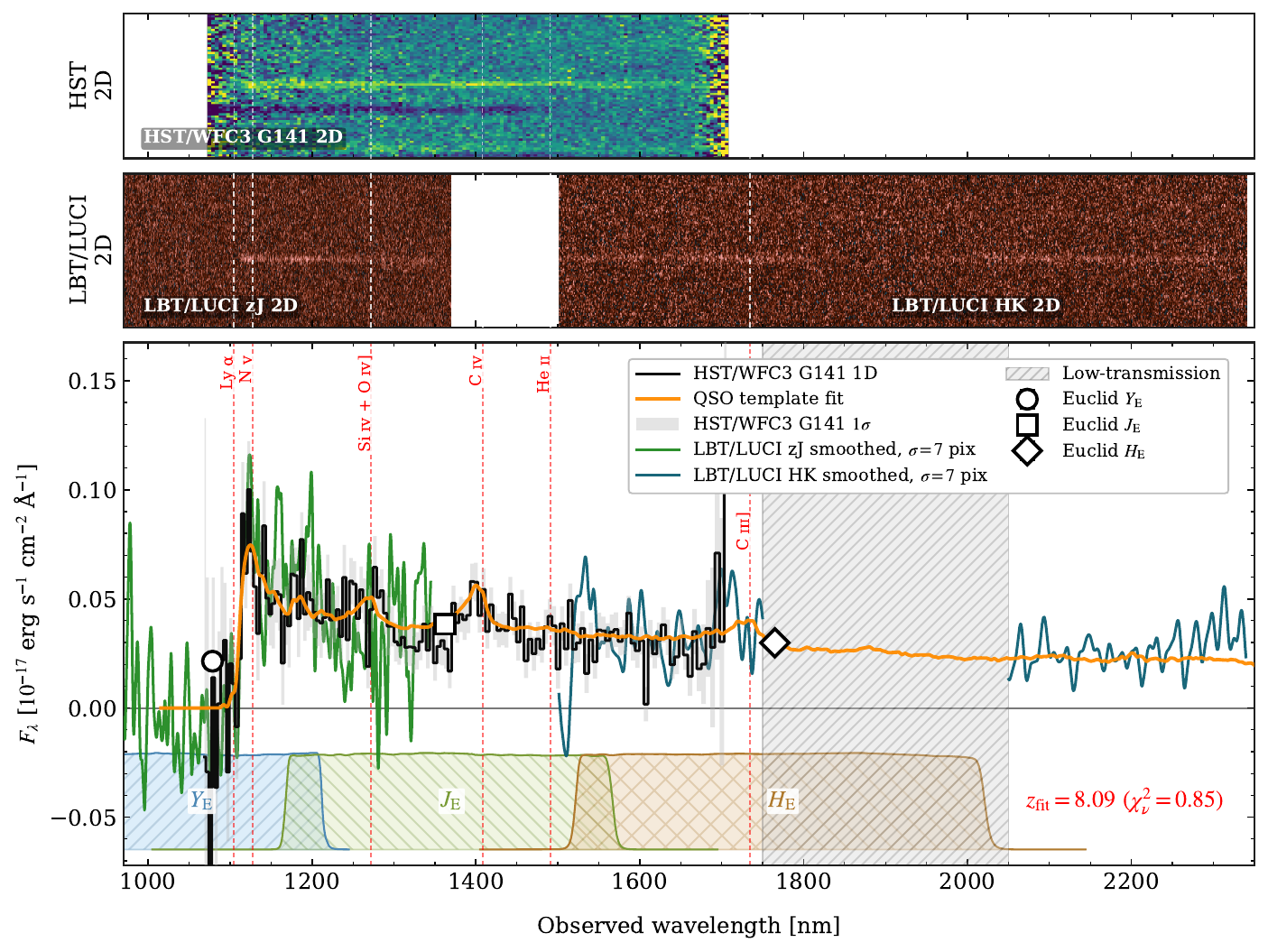}
    \caption{Combined spectroscopic confirmation of J1504+6252. The upper panels show the two-dimensional spectra from HST/WFC3 G141 and LBT/LUCI G200 zJspec and HKspec. The lower panel compares the extracted HST/WFC3 G141 spectrum with the ground-based LUCI spectra and the best-fitting quasar template, with the spectra placed on a common flux scale using the \Euclid \YE, \JE, and \HE photometry. The \Euclid photometric measurements are shown as open symbols, and the shaded bands indicate the corresponding \Euclid/NISP filter transmission curves \citep[][]{Schirmer-EP18}. The sharp continuum break at $\lambda_{\rm obs}\simeq1.1\,\micron$ is identified as the Ly$\,\alpha$ break, yielding a template redshift of $z_{\rm fit}=8.09$ with $\chi^2_\nu=0.86$.
    Grey shading shows the $1\sigma$ uncertainty on the HST/WFC3 spectrum. Vertical dashed lines mark the expected observed wavelengths of prominent rest-frame UV emission lines at the best-fitting redshift.}
    \label{fig:spec}
\end{figure*}

%Silvia's version
To independently confirm the spectral break seen near the blue edge of the HST/WFC3 G141 grism,  and to extend the wavelength coverage, we obtained deep NIR long-slit spectroscopy of J1504+6252 with LBT/LUCI. The observations were carried out in binocular mode on 2026-05-15 and 2026-05-16 (programme ID: MPIA-2026A-001, PI E.~Ba\~nados). 
The observational setup on May 15 (16) used the G200 grating coupled with the HKspec (zJspec) filter on both the LUCI1 and LUCI2 arms. 
LUCI G200 with the zJ filter covers 900--1200\,nm (with a central wavelength of $1170$ nm), while the HK filter covers 1500--2400\,nm (centred at $1930$ nm). 
The zJspec setup is therefore expected to cover the Ly$\,\alpha$ break, while the HKspec setup is expected to detect the \ciii emission line.
We employed the $1\arcsec$ slit at a position angle of \ang{280.83;;} and utilised an ABBA dithering pattern for optimal sky subtraction. Individual science exposures were 240\,s in both setups. Each setup consisted of 51 science exposures per LUCI arm, corresponding to 12\,240\,s per arm.  However, due to poor seeing conditions during the two observing nights (with seeing degrading to 1\farcs8), not all exposures were used in producing the final spectrum shown in Fig.~\ref{fig:spec}.
For the zJspec setup, 49 exposures per arm were retained, yielding 98 individual exposures combining LUCI1 and LUCI2, for a total exposure time of 23\,520\,s. For the HKspec setup, 40 exposures from LUCI2 and 49 exposures from LUCI1 were retained, yielding a total of 89 exposures and a total exposure time of 21\,360\,s.

The LUCI spectra were reduced with \texttt{PypeIt} (version 1.16.1) using standard NIR long-slit reduction procedures \citep[following e.g.][]{onorato2025,belladitta2025}. These include bias subtraction, flat fielding, sky subtraction, wavelength calibration using exposures of arc lamps, flux calibration using exposures of spectrophotometric standard stars, and telluric correction.
The individual exposures were reduced separately, then combined with the \texttt{coadd2d} routine, and optimal extraction was applied to obtain the 1D spectrum. The final spectrum was flux calibrated to match the \Euclid NIR photometry.

The resulting 2D and 1D LUCI spectra shown in Fig.~\ref{fig:spec} independently confirm the presence of the spectral break at $\lambda_{\rm obs}\simeq{\rm 1120}\,{\rm nm}$ inferred from the HST/WFC3 data, with continuum emission detected redwards of the break. From the scaled LBT/LUCI zJ and HK spectroscopy, we measure a rest-frame UV continuum spectral index of $\alpha_\lambda=-1.28$ ($f_\lambda\propto\lambda^{\alpha_\lambda}$). This relatively flat continuum is broadly consistent with typical quasar SEDs \citep[$\alpha_\lambda\simeq -1.5$, e.g.][]{vandenberk2001}.

\subsection{JWST/NIRSpec fixed slit spectroscopy}
\label{sec:jwst}

We subsequently performed follow-up restframe UV to optical spectroscopy of J1504+6252 with JWST. The observations were conducted on 2026-06-26 with NIRSpec in fixed-slit mode through the Director’s Discretionary Discovery programme GO 12761\footnote{\url{https://www.stsci.edu/jwst/science-execution/program-information?id=12761}} (PI: J.~Wolf).

The target was observed in fixed-slit mode using the G140H/F100LP and G235H/F170LP high-resolution ($R=2700$) configurations, as well as the G395M/F290LP medium-resolution ($R=1000$) configuration. Target acquisition was performed with the F140X filter using the NRSRAPID readout pattern. The G140H and G235H observations used the S200A1/S200A2 fixed-slit configuration, the full detector subarray, the NRSIRS2RAPID readout pattern, one integration, and 56 and 42 groups, respectively. The G395M observation used the S200A2 fixed slit with the full detector subarray, NRSIRS2RAPID readout, one integration, and 56 groups. The total on-source integration times in the G140H/F100LP, G235H/F170LP, and G395M/F290LP gratings were 788\,s, 591\,s, and 788\,s, respectively.
Together, the three grating settings provide continuous coverage from the \lya break through \civshort, \ciii, \mgii (hereafter \mgiishort), \hbetashort, the \oiiidd doublet (hereafter \oiiishort) and the rest-frame UV/optical continuum, enabling a precise systemic redshift estimate and measurements of the quasar continuum and broad emission-line properties.

The fixed-slit exposures were processed from the uncalibrated data products using the JWST Calibration Pipeline (v1.20.2) with CRDS context map \texttt{jwst\_1464.pmap}. Stages~1 and~2 (\texttt{calwebb\_detector1} and \texttt{calwebb\_spec2}) applied the standard detector-level corrections, assigned the wavelength solution, and performed the flat-field, point-source path-loss, and flux calibration steps using the CRDS reference files. We independently verified the absolute spectrophotometric calibration by computing synthetic Euclid measurements from the flux-calibrated NIRSpec spectrum. We obtain \YE$=24.00$, \JE$=22.94$, and \HE$=22.70$ mag, in good agreement with the measured Euclid photometry of \YE$=24.10\pm0.16$, \JE$=22.96\pm0.06$, and \HE$=22.67\pm0.06$ mag. The synthetic photometry agrees with the Euclid measurements to within $0.1$ mag in all three bands, corresponding to $\sim10\%$ in \YE and better than $3\%$ in \JE and \HE. We disabled the background subtraction, \texttt{resample\_spec}, and \texttt{extract\_1d} steps, and instead passed the calibrated two-dimensional spectral images to \texttt{PypeIt} (v1.17.3). Sky subtraction was performed using \texttt{PypeIt}'s local sky-subtraction procedure, which jointly models the source profile and sky spectrum in each exposure. One-dimensional spectra were then optimally extracted for each exposure and grating setting. The individual spectra were placed on a common wavelength grid, scaled to a consistent flux level, and coadded with \texttt{PypeIt} to produce the final NIRSpec spectrum used for the redshift and emission-line analysis.

\begin{figure*}
    \centering
    \includegraphics[width=1.0\linewidth]{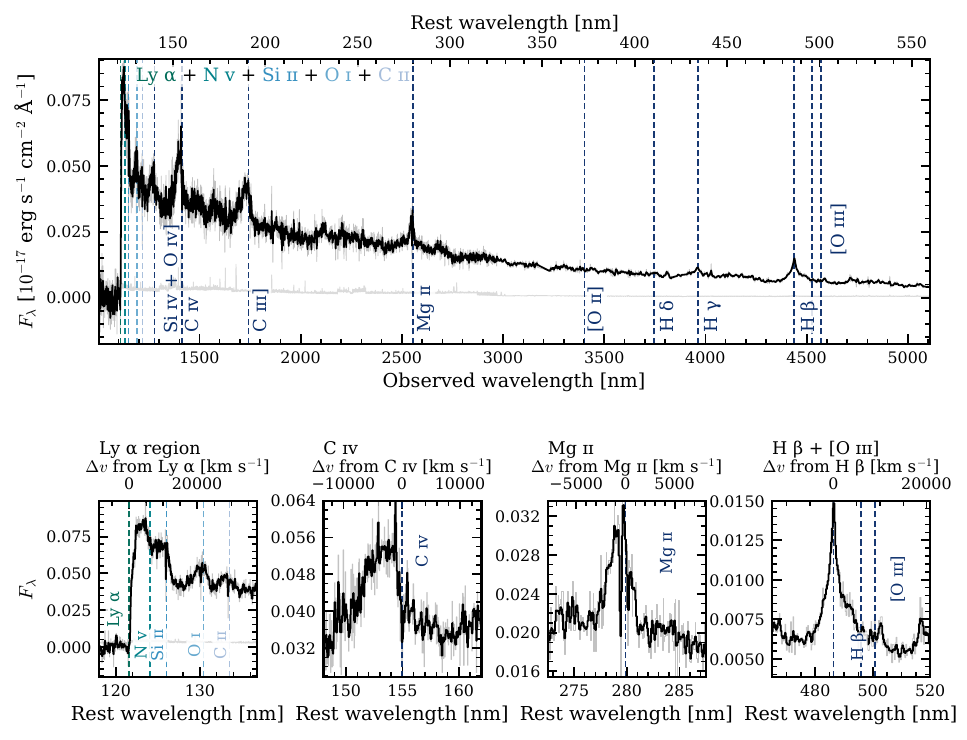}
    \caption{JWST/NIRSpec spectrum of J1504+6252. The upper panel shows the combined fixed-slit NIRSpec spectrum in the observed frame, stitched together from the G140H, G235H, and G395M gratings, with the rest-frame wavelength axis shown at the top for the adopted redshift $z=8.129$ (see Sect.~\ref{sec:bhm}). The black line shows the extracted one-dimensional spectrum and the grey curve shows the corresponding flux uncertainty. Vertical dashed lines mark the expected wavelengths of prominent rest-frame UV and optical emission lines. The lower panels show zoom-ins on the \lya+\nv, \civshort, \mgiishort, and \hbetashort+\oiiishort regions. The NIRSpec spectrum reveals a sharp \lya break, broad UV emission lines, and broad rest-frame optical \hbetashort emission, confirming J1504+6252 as a luminous broad-line quasar at $z>8$. Additionally, an \mgiishort metal absorption feature is clearly detected close to the systemic redshift. The \lya line profile shows substantial damping. The centroid of the broad \civshort emission line is shifted by about $\sim 3500 \, \mathrm{km\, s^{-1}}$, suggesting the presence of strong nuclear outflows.}
    \label{fig:spec_jwst}
\end{figure*}

The combined spectrum from the three NIRSpec gratings is presented in Fig.~\ref{fig:spec_jwst}. It unambiguously confirms the quasar nature of J1504+6252, showing a sharp \lya break together with broad rest-frame UV and optical emission lines, including \nv, \civshort, \ciii, \mgiishort, \hbetashort. These features firmly establish J1504+6252 as a quasar. In Appendix~\ref{sec:cousins}, we present  lower-redshift spectral analogues of J1504+6252 identified from Sloan Digital Sky Survey (SDSS) Data Release~ 14 (DR14) spectroscopy using a nearest-neighbour search in spectral space. 

The spectrum also displays several distinctive features, including weak or undetected narrow \oiiishort\ emission. This characteristic, recently observed in both luminous $z>7.5$ quasars and JWST AGN \citep[e.g.][]{wolf26,caputi2026,maiolino2026}, may reflect a combination of low metallicity and collisional de-excitation at high gas densities in the nuclear region, although the absence of a clearly detected auroral \oiiiaur line challenges the high-density scenario \citep[e.g.][]{maiolino2026,ivey2026}.   
A luminosity-dependent suppression may also contribute through the narrow-line Baldwin effect observed in Type~1 AGN, whereby the strength of \oiiishort relative to the AGN continuum decreases towards higher luminosities \citep{netzer2004,stern2012}. 

The spectrum also reveals a strongly blueshifted and asymmetric broad \civshort\ emission-line profile \citep[e.g.][]{meyer2019,schindler2020}. We estimate a velocity offset of $\sim 3500\,\mathrm{km\,s^{-1}}$ for the centroid of the broad component, indicative of nuclear outflows. A detailed analysis of the outflow and broad-line region properties will be presented in a forthcoming publication (Guarneri et al., in prep.). Finally, the quasar appears to be affected by a proximate damped \lya system \citep[e.g.][]{russell2006,prochaska2008,hennawi2009,ellison2010}, as indicated by the strongly damped \lya profile and the associated metal absorption features (in particular \mgiishort) near the systemic redshift \citep[e.g.][]{rowlands2026}. These absorption features will be investigated in detail in a separate publication (Yang et al., in prep.).
In the following section, we analyse the \hbetashort--\oiiishort and \mgiishort regions to determine the definitive systemic spectroscopic redshift and measure the accretion properties from the broad emission lines.

\section{Results}
\label{sec:bhm}

\subsection{Systemic redshift and black hole mass}
We model the \hbetashort--\oiiishort region of the JWST/NIRSpec G395M spectrum with \texttt{unite} (Unified liNe Integration Turbo Engine), a Bayesian emission-line fitting framework\footnote{\url{https://github.com/TheSkyentist/unite/}} \citep[][]{hviding_2026_20734477}. \texttt{unite} forward-models the observed spectrum by combining continuum emission, line profiles, and the wavelength-dependent instrumental resolution, and samples the posterior distribution of the model parameters with a Hamiltonian Monte Carlo approach. The fit is performed over the observed-frame wavelength range 4000--4800\,nm.

The spectral model comprises a power-law continuum, the optical \feii template of \citet{Park2022}, the \hbetashort emission complex, and the \oiiishort doublet. The \feii template amplitude and intrinsic velocity width are fitted independently, with a uniform prior of 1000--15\,000\,km\,s$^{-1}$ on its FWHM \citep[e.g.][]{hu08}. To determine the level of complexity required by the \hbetashort profile, we compare three nested models: a single broad component; a narrow core; plus a broad component, and a narrow core plus intermediate- and broad-width components. We adopt uniform intrinsic FWHM priors of 30--700\,km\,s$^{-1}$, 700--2500\,km\,s$^{-1}$, and 2500--15\,000\,km\,s$^{-1}$ for the narrow, intermediate, and broad components, respectively. The narrow \hbetashort component and the \oiiishort doublet are tied to a common redshift and velocity width, whereas the intermediate and broad \hbetashort components are allowed independent velocity offsets and widths. The \oiiishort doublet ratio is fixed to its atomic value, \(F_{5007}/F_{4959}=2.98\), and its flux is constrained to be non-negative so that, in the absence of a significant detection, the posterior yields a one-sided upper limit.

The preferred \hbetashort decomposition is selected using posterior-predictive model comparison where available, with the Bayesian information criterion used as a fallback. In either case, we retain the simplest model whose performance is statistically indistinguishable from that of the best-fitting alternative, and require any additional component to have a constrained, non-zero flux posterior. The model also includes a multiplicative scaling parameter for the NIRSpec/G395M line-spread function. All reported line widths are intrinsic because the model profiles are forward-convolved with the instrumental line-spread function \citep{degraaff24} before comparison with the data. We present the result of this fit in Fig.~\ref{fig:fit_hb}. The detailed emission line properties of the decomposition are listed in Appendix \ref{sec:appendix}. The model comparison strongly favours the three-component \hbetashort\ decomposition, comprising narrow, intermediate, and broad components, with $\Delta{\rm BIC}=28.8$ relative to the two-component model.

The \oiiishort\ doublet is not formally detected in the fit. We therefore quote a posterior $3\sigma$ upper limit of $L_{\mathrm{[O_{III}]\lambda 5007}}<5.7\times10^{41}\,{\rm erg\,s^{-1}}$. We measure the relative strength of \feii\ as $R_{\rm FeII}=F_{\rm FeII}(4434$--$4684\,\text{\AA})/
F_{\rm H\,\beta,BLR}$, where the \hbetashort\ BLR flux includes the
intermediate and broad components, and obtain
$R_{\rm FeII}=0.77^{+0.06}_{-0.05}$. The independently fitted optical \feii\ template requires an additional Gaussian convolution with a ${\rm FWHM}$ of $1270^{+190}_{-150}\,{\rm km\,s^{-1}}$.\footnote{Because the empirical template already contains finite intrinsic line broadening, this quantity represents only the additional Gaussian convolution applied to the template and not the intrinsic width of the \feii\ emission.}

We also note that we do not detect the host-galaxy \oii emission line, which could, in principle, provide an upper limit on the unobscured star-formation rate of the host galaxy \citep[e.g.][]{Feng26}; however, a quantitative assessment is beyond the scope of this discovery paper.

Because the \oiiishort doublet is not significantly detected, we determine the systemic redshift of J1504+6252 from the fitted narrow \hbetashort component \citep[e.g.][]{bon2020}. The \oiiishort doublet would generally be preferred as a systemic-redshift tracer because it arises in the narrow-line region and is less affected by broad-line-region kinematics, although its centroid can itself be shifted by ionised-gas outflows. Other available tracers, such as \mgiishort, are broad emission lines and therefore provide less precise constraints on the systemic velocity. The only significantly detected narrow lines in the G395M spectrum are \hbetashort and \hgamma, and narrow \hbetashort therefore provides the best available rest-frame optical systemic-redshift indicator. Large-sample SDSS studies show that narrow Balmer lines generally remain close to the host-galaxy stellar velocity, supporting their use as systemic-redshift tracers when stellar absorption features are unavailable \citep{bae2014}.

We measure
$z_{\rm sys,nar}=8.129\pm0.005_{\rm tot}$ (where the total uncertainty accounts for
$\pm0.001_{\rm stat}$ statistical uncertainty and $\pm0.005_{\rm sys}$ systematic uncertainty). The statistical and systematic components correspond to velocity uncertainties of approximately 33 and $150\,{\rm km\,s^{-1}}$, respectively, with the total uncertainty dominated by the systematic term. The statistical uncertainty reflects the formal uncertainty on the fitted narrow-\hbetashort centroid. Although \citet{bon2020} found that the narrow components of \hbetashort and \oiiishort were consistent with the mean stellar velocity to within $60\,{\rm km\,s^{-1}}$ in lower-redshift quasars, \citet{bian2005} measured an object-to-object dispersion of $158\,{\rm km\,s^{-1}}$ between the narrow-\hbetashort and \oiiishort centroids in SDSS narrow-line Seyfert~1 galaxies, including a tail of strongly blueshifted \oiiishort sources. We therefore conservatively adopt a systematic uncertainty of $150\,{\rm km\,s^{-1}}$ to account for narrow-line-region kinematics and for the uncertain applicability of low-redshift calibrations to a luminous quasar at $z>8$.

The intrinsic FWHM of the narrow component is
$
430^{+150}_{-110}\,{\rm km\,s^{-1}}$.
The intermediate and broad \hbetashort\ components have intrinsic FWHMs of
$
2290^{+140}_{-220}\,{\rm km\,s^{-1}}$
and
$
7800^{+530}_{-560}\,{\rm km\,s^{-1}}$,
respectively. For the virial analysis, we define the broad-line-region
profile as the sum of the intermediate and broad components, excluding
the narrow core, and measure a FWHM of $
3160^{+270}_{-270}\,{\rm km\,s^{-1}}$.

We convert the measured \hbetashort\ line width into black hole mass estimates using several calibrations that combine a luminosity-based proxy for the broad-line-region radius \citep[][]{Kaspi2000,Bentz2013}, with the broad-line velocity width hereafter referred to as single-epoch calibrations \citep[e.g.][]{Greene2005,Vestergaard2006,shen2011}.
 As our benchmark, we adopt the widely used continuum-based prescription of \citet{Vestergaard2006}, facilitating direct comparison with previous studies of luminous high-redshift quasars \citep[e.g.][]{Liu24,loiacono24}:
\begin{equation}
\begin{aligned}
\logten\left(\frac{M_{\rm BH}}{M_\odot}\right)
={}& 6.91
+ 0.50\logten\left[
\frac{\lambda L_\lambda(5100\,\text{\AA})}
{10^{44}\,\mathrm{erg\,s^{-1}}}
\right]
\\
&+ 2\logten\left[
\frac{{\rm FWHM}({\mathrm{H\,\beta,BLR})}}
{10^3\,\mathrm{km\,s^{-1}}}
\right]\, .
\end{aligned}
\end{equation}

Here, $\lambda L_\lambda(5100\,\text{\AA})$ is the featureless continuum luminosity at rest-frame $5100\,\text{\AA}$ (see Sect.~\ref{sec:lumin}), excluding the \feii contribution. Applying the relation to the posterior samples of the continuum luminosity and total BLR profile, and including the intrinsic scatter of the calibration and the adopted object-to-object virial-factor uncertainty, we obtain
\begin{equation}
\logten\left(\frac{M_{\rm BH}}{M_\odot}\right)=
8.56^{+0.07}_{-0.08}\,(\mathrm{stat})
^{+0.43}_{-0.43}\,(\mathrm{tot})\, ,
\end{equation}

\noindent where the first quoted asymmetric uncertainty is statistical and the second one is
the total uncertainty including systematics. We complement this benchmark estimate with two alternative \hbetashort\ prescriptions. The recent FWHM-based fundamental-plane calibration of \citet{woo2026}, applied to the same total BLR profile, yields $\logten(M_{\rm BH}/M_\odot)=8.33^{+0.38}_{-0.38}$. The broad  \hbetashort\ luminosity calibration of \citet{dallabonta2020} instead gives $\logten(M_{\rm BH}/M_\odot)=8.89^{+0.40}_{-0.40}$. Our \hbetashort-based mass estimates rely on complementary calibrations: \citet{Vestergaard2006} use the optical continuum luminosity, whereas \citet{dallabonta2020} use the broad \hbetashort\ luminosity as a proxy for the BLR radius, thereby reducing sensitivity to host-galaxy contamination. Finally, \citet{woo2026} include the Eddington ratio as an additional parameter in their calibration, accounting empirically for accretion-rate-dependent departures from the standard radius--luminosity relation.

As an independent consistency check, we also estimate the black hole mass from the \mgiishort\ doublet covered by the G235H grating. With our \texttt{unite} framework, we fit the \mgiishort\ region over the observed-frame wavelength range $2300$--$2900$\,nm, masking absorption features that overlap the emission-line profile and would otherwise bias its inferred shape and width. In this spectral window, we use the \feii\ template of \citet{salviander2007}, who reconstruct the emission beneath \mgiishort\ from the \citet{vestergaard2001} template using the theoretical \feii\ model of \citet{sigut2003}. As for the \hbetashort\ region, we compare models containing one, two, and three kinematic components and select the simplest adequate model. The single broad-component model is preferred, with an intrinsic FWHM of $3740^{+160}_{-170}\,{\rm km\,s^{-1}}$.

We first apply the established \mgiishort\ calibration of \citet{vestergaardosmer2009}, obtaining
$\logten(M_{\rm BH}/M_\odot)=8.76^{+0.55}_{-0.55}$.
For comparison, the \mgiishort\ fundamental-plane calibration of \citet{woo2026} yields
$\logten(M_{\rm BH}/M_\odot)=8.50^{+0.39}_{-0.39}$. The fit to the \mgiishort\ region is shown in the lower panel of Fig.~\ref{fig:fit_hb}.

All \hbetashort- and \mgiishort-based estimates are mutually consistent within their uncertainties and are summarised in  Table~\ref{tab:j1504_properties}. We quote the \citet{Vestergaard2006} \hbetashort\ result as our benchmark single-epoch mass estimate, while recognising that the dominant uncertainty arises from the adopted calibration. The extreme central values span

\begin{equation}
\logten\left(\frac{M_{\rm BH}}{M_\odot}\right)=8.33\text{--}8.89\,,
\end{equation}

\noindent which we use to represent the calibration-dependent range of plausible black hole masses for J1504+6252.

\begin{figure}
    \centering
    \includegraphics[]{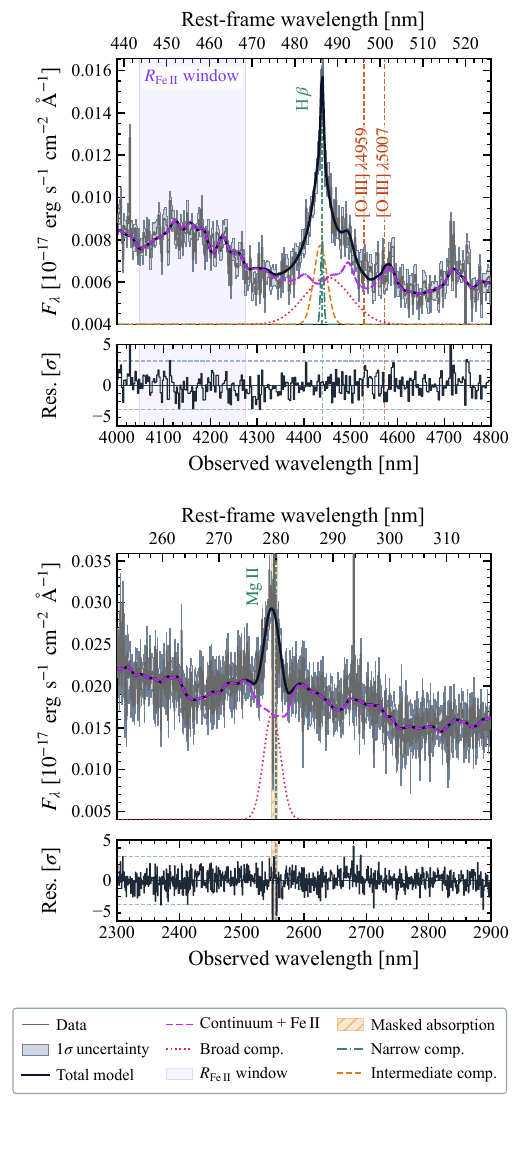}
    \caption{
Fit to the \hbetashort--\oiiishort\ (top) and \mgiishort\ (bottom) regions of J1504+6252. For each region, the upper panel shows the JWST/NIRSpec spectral fit. %, with G395M used for the \hbetashort--\oiiishort\ region and G235M used for the \mgiishort\ region. 
The observed spectrum is shown as the dark grey histogram, and the $1\sigma$ flux uncertainty is indicated in light grey. The best-fitting total model is shown as the solid black curve, while the continuum is shown as the dashed magenta curve. Individual emission-line model components are shown as coloured offset curves. In the \hbetashort panel, the shaded purple region marks the rest-frame wavelength interval used to measure $R_{\rm FeII}$ from the \feii template component. In the \mgiishort\ panel, the hatched orange region marks the automatically masked narrow absorption region excluded from the fit. Vertical dashed lines mark the expected observed wavelengths of systemic narrow \hbetashort, \oiiishort, and \mgiishort. No significant \oiiishort\ emission is detected. The lower panels show the residuals in units of the flux uncertainty. The horizontal lines in the residual panel delimit the 3$\sigma$ range.}
    \label{fig:fit_hb}
\end{figure}

\subsection{Continuum luminosity and Eddington rate}
\label{sec:lumin}

We measure the UV magnitude at 145 nm, $M_{1450}$, from a sigma-clipped, inverse-variance-weighted linear continuum fit over $142.5$--$147.5\,\mathrm{nm}$ in the rest frame, corresponding to the G235H portion of the JWST/NIRSpec spectrum. For all luminosity measurements, we include a 5\% systematic uncertainty to account for the absolute flux calibration accuracy of the NIRSpec fixed-slit spectroscopy quoted by the JWST documentation at the time of writing\footnote{\url{https://jwst-docs.stsci.edu/jwst-calibration-status/nirspec-calibration-status/}}. We obtain $F_\lambda(145\,\mathrm{nm})=(3.47\pm0.18)\times10^{-19}\,\mathrm{erg\,s^{-1}\,cm^{-2}\,\si{\angstrom}^{-1}}$, corresponding to $M_{1450}=-24.08\pm0.06$.
We then measure the monochromatic continuum luminosity at rest-frame 510\,nm\ from the best-fitting \hbetashort--\oiiishort\ spectral model, obtaining $\logten\left[\lambda L_{\lambda}(5100\,\si{\angstrom})/({\rm erg\,s^{-1}})\right]=45.30\pm0.02\,$.

We convert the monochromatic quantity to a bolometric luminosity using the
luminosity-dependent correction of \citet{runnoe2012}, accounting for the correction presented by \citet{runnoe12erratum}. Here, $L_{\rm iso}$ is the
isotropic-equivalent bolometric luminosity inferred from the observed continuum:
\begin{equation}
\logten\left[
\frac{L_{\rm iso}}{\mathrm{erg\,s^{-1}}}
\right]
=
4.891
+
0.912\,\logten\left[
\frac{\lambda L_\lambda(5100\,\si{\angstrom})}
{\mathrm{erg\,s^{-1}}}
\right]\,.
\end{equation}
We then adopt $L_{\rm bol}=0.75\,L_{\rm iso}$, which results in
\begin{equation}
\logten\left[
\frac{L_{\rm bol}}{\mathrm{erg\,s^{-1}}}
\right]
=
46.08 \pm 0.02\,.
\end{equation}

\begin{figure}
    \centering
    \includegraphics[width=1.0\linewidth]{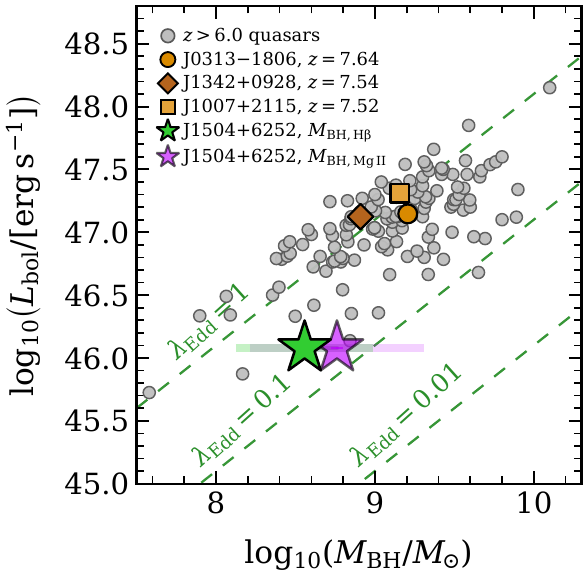}
    \caption{$L_{\rm bol}$ versus $M_{\rm BH}$ for quasars $z>6.0$ from the literature \citep[][and references therein]{belladitta2025} and J1504+6252. Brown symbols mark the previously most distant quasars with black hole mass measurements \citep{banados2018a,Yang2020,wang2021}. The two stars show the \hbetashort and \mgiishort based  $M_\mathrm{BH}$ measurements for  J1504+6252, using the calibrations from \citet{Vestergaard2006} and \citet{vestergaardosmer2009}, respectively. Dashed lines indicate constant Eddington ratios. J1504+6252 is offset in terms of its Eddington rate with respect to the bulk of the known $z>6$ quasar population.}
    \label{fig:mbhlbol}
\end{figure}

Combining this bolometric luminosity with the black hole mass estimates presented in Sect.~\ref{sec:bhm}, we compute the Eddington ratio as $\lambda_{\rm Edd}=L_{\rm bol}/L_{\rm Edd}$, where $L_{\rm Edd}=1.26\times10^{38} \, (M_{\rm BH}/M_\odot)\, {\rm erg\,s^{-1}}$ \citep[e.g.][]{netzer2013}. Across the adopted \hbetashort\ and \mgiishort\ single-epoch calibrations, we obtain central values spanning

\begin{equation}
\lambda_{\rm Edd}=0.12\text{--}0.44\, .
\end{equation}

\noindent  with the individual measurements and uncertainties listed in Table~\ref{tab:j1504_properties}.
Thus, J1504+6252 is inferred to be accreting at approximately $12$--$44$\% of the Eddington limit, depending on the adopted black hole mass calibration. These values place it towards the lower envelope of the $z>6$ quasar population, although such values are not unprecedented, as shown in Fig.~\ref{fig:mbhlbol}. 

\begin{table}
\caption{Summary of the observed and derived properties of J1504+6252.}
\label{tab:j1504_properties}
\centering
\begingroup
\renewcommand{\arraystretch}{1.16}
\begin{tabular}{lll}
\toprule
\toprule
Property & Value & \\
\midrule
\multicolumn{3}{l}{Source properties} \\
\addlinespace[0.5pt]

ID & \multicolumn{2}{l}{EUCL\,J150452.291+625224.39} \\[-1.5pt]
Name & \multicolumn{2}{l}{Dún Aonghasa} \\[-1.5pt]
RA (J2000) & \multicolumn{2}{l}{$15{:}04{:}52.291$} \\[-1.5pt]
Dec (J2000) & \multicolumn{2}{l}{$+62{:}52{:}24.39$} \\[-1.5pt]
$z_{\rm sys,nar}$ &
\multicolumn{2}{l}{$8.129 \pm 0.001_{\rm stat} \pm 0.005_{\rm sys}$} \\[-1.5pt]
$M_{1450}$ [AB mag] &
\multicolumn{2}{l}{$-24.08 \pm 0.06$} \\

\midrule
\multicolumn{3}{l}{\Euclid Photometry} \\
\addlinespace[1.5pt]
Band &
AB magnitude &
$f_{\nu}$ [$\mu\mathrm{Jy}$] \\
\midrule

\IE &
$>26.2$ &
$-0.009 \pm 0.009$ \\[-1pt]

\YE &
$24.10 \pm 0.16$ &
$0.833 \pm 0.124$ \\[-1pt]

\JE &
$22.96 \pm 0.06$ &
$2.385 \pm 0.136$ \\[-1pt]

\HE &
$22.67 \pm 0.06$ &
$3.109 \pm 0.169$ \\

\midrule
\multicolumn{3}{l}{Emission-line and continuum properties} \\
\addlinespace[2.5pt]

FWHM(\hbetashort,BLR) &
\multicolumn{2}{l}{$3160^{+270}_{-270}\,\mathrm{km\,s^{-1}}$} \\
\addlinespace[1pt]

FWHM(\mgiishort,BLR) &
\multicolumn{2}{l}{$3740^{+160}_{-170}\,\mathrm{km\,s^{-1}}$} \\
\addlinespace[1pt]

$R_{\rm FeII}$ &
\multicolumn{2}{l}{$0.77^{+0.06}_{-0.05}$} \\
\addlinespace[1pt]

$\logten L_{\rm H\,\beta,BLR}$ &
\multicolumn{2}{l}{$43.52^{+0.03}_{-0.03}\,\mathrm{erg\,s^{-1}}$} \\
\addlinespace[1pt]

$\logten \lambda L_{\lambda}(3000\,\si{\angstrom})$ &
\multicolumn{2}{l}{$45.51^{+0.02}_{-0.02}\,\mathrm{erg\,s^{-1}}$} \\
\addlinespace[1pt]

$\logten \lambda L_{\lambda}(5100\,\si{\angstrom})$ &
\multicolumn{2}{l}{$45.30^{+0.02}_{-0.02}\,\mathrm{erg\,s^{-1}}$} \\
\addlinespace[1pt]

$\logten L_{\rm bol}$ &
\multicolumn{2}{l}{$46.08^{+0.02}_{-0.02}\,\mathrm{erg\,s^{-1}}$} \\

\midrule
Single-epoch cal. &
$\logten(M_{\rm BH}/M_\odot)$ &
$\lambda_{\rm Edd}$ \\
\midrule
\multicolumn{3}{l}{\hbetashort-based} \\
\addlinespace[2pt]

W26 &
$8.33^{+0.10\,(+0.38)}_{-0.11\,(-0.38)}$ &
$0.44^{+0.12\,(+0.62)}_{-0.09\,(-0.26)}$ \\
\addlinespace[1.5pt]

DB20 &
$8.89^{+0.05\,(+0.40)}_{-0.06\,(-0.40)}$ &
$0.12^{+0.02\,(+0.19)}_{-0.01\,(-0.08)}$ \\
\addlinespace[1.5pt]

VP06 &
$8.56^{+0.07\,(+0.43)}_{-0.08\,(-0.43)}$ &
$0.26^{+0.05\,(+0.45)}_{-0.04\,(-0.17)}$ \\

\midrule
\multicolumn{3}{l}{\mgiishort-based} \\
\addlinespace[2pt]

W26 &
$8.50^{+0.05\,(+0.39)}_{-0.06\,(-0.39)}$ &
$0.30^{+0.04\,(+0.44)}_{-0.04\,(-0.18)}$ \\
\addlinespace[1.5pt]

VO09 &
$8.76^{+0.04\,(+0.55)}_{-0.04\,(-0.55)}$ &
$0.17^{+0.02\,(+0.42)}_{-0.02\,(-0.12)}$ \\

\bottomrule
\end{tabular}
\endgroup

\tablefoot{
The coordinates correspond to the centroid of the \Euclid NISP \JE-band
detection of J1504+6252, as determined by the \Euclid NIR
catalogue-extraction pipeline \citep{Q1-TP003}. The \Euclid photometry is based on PSF flux measurements. The systemic redshift is measured from the narrow \hbetashort{} component. The FWHM of the \hbetashort BLR component  is measured from the sum of the intermediate and broad components. The FWHM of the  \mgiishort BLR component is measured from the broad-line profile. The bolometric luminosity is computed from $\lambda L_{\lambda}(5100\,\si{\angstrom})$ using the luminosity-dependent bolometric correction of \citet{runnoe2012}, with the coefficients provided by \citet{runnoe12erratum}. The quoted uncertainty on $L_{\rm bol}$ propagates only the uncertainty in $\lambda L_{\lambda}(5100\,\si{\angstrom})$ through the fixed central bolometric correction. The uncertainties in the \citet{runnoe2012} coefficients are not included because the covariance between the fitted intercept and slope is unavailable. W26, DB20, VP06, and VO09 refer to \citet{woo2026}, \citet{dallabonta2020}, \citet{Vestergaard2006}, and \citet{vestergaardosmer2009}, respectively. For $M_{\rm BH}$ and $\lambda_{\rm Edd}$, the first uncertainty is statistical, while the value in parentheses is the total uncertainty including the intrinsic scatter and adopted systematic uncertainty of the corresponding single-epoch mass calibration. The total $\lambda_{\rm Edd}$ uncertainties do not include an additional uncertainty from the \citet{runnoe2012} bolometric-correction calibration.
}
\end{table}

\section{A new quasar redshift frontier: Implications for
SMBH growth}
\label{sec:disc}

\begin{figure*}
    \sidecaption
    \includegraphics[width=12cm]{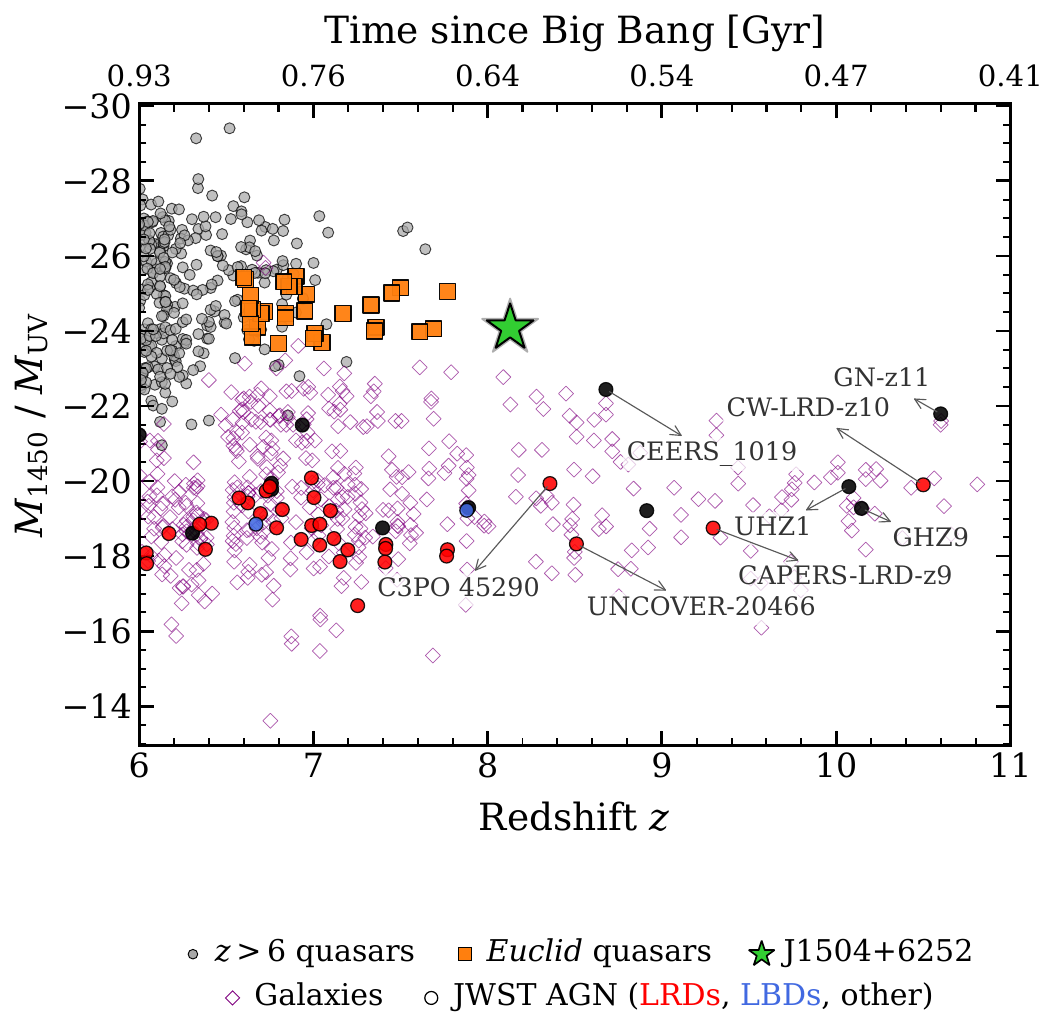}
    \caption{Absolute UV magnitude, $M_{1450}$ or $M_{\rm UV}$, as a function of redshift for high-redshift quasars, galaxies, LRDs, LBDs, and other JWST-selected AGN up to $z=11$. Previously known quasars are shown as grey circles \citep[][and references therein, as well as quasars discovered by \citealt{yang23}, \citealt{wolf24}, \citealt{belladitta2025}, and \citealt{martinez2026}]{Fan2023}. Orange squares show recently identified \Euclid quasars from \citet{yang2026}. Lyman-break galaxies are shown as open purple diamonds; see \citet{Harikane2025}, \citet{SmitBowler2026}, and references therein. JWST-selected AGN are shown as circles and colour-coded according to their phenomenological classification: LRDs in red, LBDs in blue, and other AGN in black. The latter include sources from \citet{Larson2023}, \citet{kokorev2023}, \citet{harikane2023}, \citet{maiolino2024}, \citet{maiolino24}, \citet{Napolitano2025}, and \citet{Juodzbalis2026}. The LRD sample is drawn from \citet{degraaff2026}, \citet[][including the spectroscopic confirmation of CW-LRD-z10 at $z>10$ from T. Tanaka, pers. comm.]{Tanaka2025}, and \citet{geris2026}, while LBDs are taken from \citet{geris2026}. The green star marks J1504+6252, which is more UV-luminous than the LRDs, other JWST-selected AGN, and Lyman-break galaxies at comparable redshifts.}
    \label{fig:zmuv}
\end{figure*}

In Fig.~\ref{fig:zmuv}, we compare J1504+6252 to the known high-redshift quasar population \citep[][and references therein]{Fan2023} in the $M_{\rm 1450}$--redshift plane, together with spectroscopically confirmed galaxies \citep[][and references therein]{SmitBowler2026} and JWST-selected lower-luminosity AGN \citep[][]{Larson2023,kokorev2023,harikane2023,maiolino2024,maiolino24,Napolitano2025,Juodzbalis2026}, including Little Red Dots and the more recently defined Little Blue Dots \citep[LRDs and LBDs;][]{hviding2025,degraaff2026,taylor2025,Tanaka2025,geris2026}.
 J1504+6252 extends the confirmed quasar redshift frontier beyond $z=8$. Its luminosity places it above the bulk of the known galaxy and LRD populations at similar redshifts, but within the extrapolation of the luminous quasar sequence to earlier times. 
The broad rest-frame UV and optical emission lines, high UV luminosity, and large black hole mass of J1504+6252 identify it as a luminous, unobscured quasar powered by a mature accreting SMBH. 

While evidence for massive black hole growth at higher redshift has been reported previously, those systems probe a fundamentally different population. The physical nature of LRDs, for instance, remains under active debate, with several observed systems pointing towards accreting black holes embedded in extremely dense gas environments \citep[e.g.][]{Ji2025,degraaff2025b,naidu2026}, a picture also explored in theoretical models involving compact, optically thick gaseous envelopes and super-Eddington accretion \citep[e.g.][]{inayoshi2025,kido25}, while alternative interpretations invoke no actively accreting black hole at all \citep[][]{baggen2024}. The JWST identification of compact sources with blue SEDs that have been termed LBDs has raised the possibility that these objects and LRDs may represent related manifestations of a common underlying population. In proposed unified scenarios, differences in circumnuclear obscuration, its geometry, and/or viewing orientation can produce the observed diversity in SED shape \citep{brazzini2026,madau2026}. JWST has also uncovered evidence for SMBH accretion in several galaxies at $z>8.5$ through broad emission lines \citep[e.g. CEERS\_1019,][although \citealt{Zamora2025} find evidence for outflow-driven line broadening]{Larson2023}, high-density gas tracers and typical AGN emission lines \citep[GN-z11,][]{maiolino24}, and emission-line diagnostics combined with X-ray observations \citep[GHZ9,][]{kovacs24,Napolitano2025}. UHZ1, at $z=10.073$, has been interpreted as a heavily obscured AGN associated with a lensed galaxy based on reported X-ray emission \citep{bogdan24,goulding2023}, although its AGN nature remains debated \citep{alvarez2026,zou2026}.
In contrast, J1504+6252 is the first spectroscopically confirmed luminous quasar at $z>8$ and exhibits the canonical spectral characteristics of lower-redshift quasars, as highlighted by the SDSS nearest neighbour search in Appendix \ref{sec:cousins}. In particular, its luminosity and broad emission lines enable the robust application of single-epoch black hole mass estimators calibrated using analogous quasar samples at lower redshift. J1504+6252 directly reveals an accreting SMBH with a mass of $\sim 10^{8.5}\,M_\odot$ accreting at $\sim 26\%$ of the Eddington limit. Its discovery demonstrates that by $z=8.129$, only about $630$ million years after the Big Bang, at least some SMBHs had already completed their obscured growth phase and evolved into the luminous quasar population. 

The existence of J1504+6252 at $z=8.13$ places new constraints on the
early growth of black holes. With a black hole mass of
$\logten(M_{\rm BH}/M_\odot)= 8.3$--$8.9$ at this cosmic epoch, the source leaves limited time for
growth from low-mass seeds. In particular, remnant seeds from Population III stars with
$M_{\rm seed}\sim10^2\,M_\odot$, often invoked at $z\sim20$ \citep[e.g.][]{volonteri2010,inayoshi2020}, if limited by Eddington accretion, require extremely rapid and nearly uninterrupted growth to reach the observed mass by this epoch while more massive seed channels can substantially relax the growth requirement \citep[e.g.][]{narajan2011}. Heavy seeds from the direct collapse of pristine gas clouds, emerging around $z\simeq15$--20 \citep[e.g.][]{trinca2026},
were therefore introduced to alleviate the black hole growth timescale problem \citep[e.g.][]{bromm2003,lodato2006,lodato2007,volonteri2007,habouzit16}. Alternatively, runaway collisions in dense nuclear star clusters have been presented as a possible formation channel for intermediate-mass black holes \citep[e.g.][]{devecchi2009,katz2015,sakurai2017}. Primordial black holes (PBHs) provide another possible pathway \citep{carr1974,rubin2001,carr2018}, because sufficiently massive PBHs formed in the early Universe can seed dark matter halo assembly and subsequently grow through gas accretion \citep[e.g.][]{dayal2024,zipparo2025,dayal2026}.

Beyond accretion, black-hole mergers may contribute to SMBH assembly. Recent JWST studies, however, indicate major-galaxy-merger rates of $\simeq2$--$6$ events per galaxy per Gyr at $z\gtrsim6$ \citep[e.g.][]{Duan2025,calabro2026}. Integrated over the $\sim0.45$ Gyr between $z\sim20$ and $z=8.13$, such rates correspond to an  expected $\simeq1$--$3$ major galaxy mergers along the main progenitor branch. The number of completed black-hole coalescences is likely smaller, because both progenitors must host black holes and because dynamical friction, binary hardening, and gravitational-wave inspiral introduce additional delays \citep[e.g.][]{volonteri2020}. Because mergers have an additive effect on black-hole mass, they are thus unlikely by themselves to assemble a $10^{8.3}$--$10^{8.9}\,M_\odot$ black hole by $z=8.13$ unless the progenitors were already massive. Consistently, \citet{pacucci2020} predict that accretion generally dominates the growth of $M_{\rm BH} >10^8\,M_\odot$ black holes at $z>6$. 

\begin{figure}
    \centering
    \includegraphics[width=1.0\linewidth]{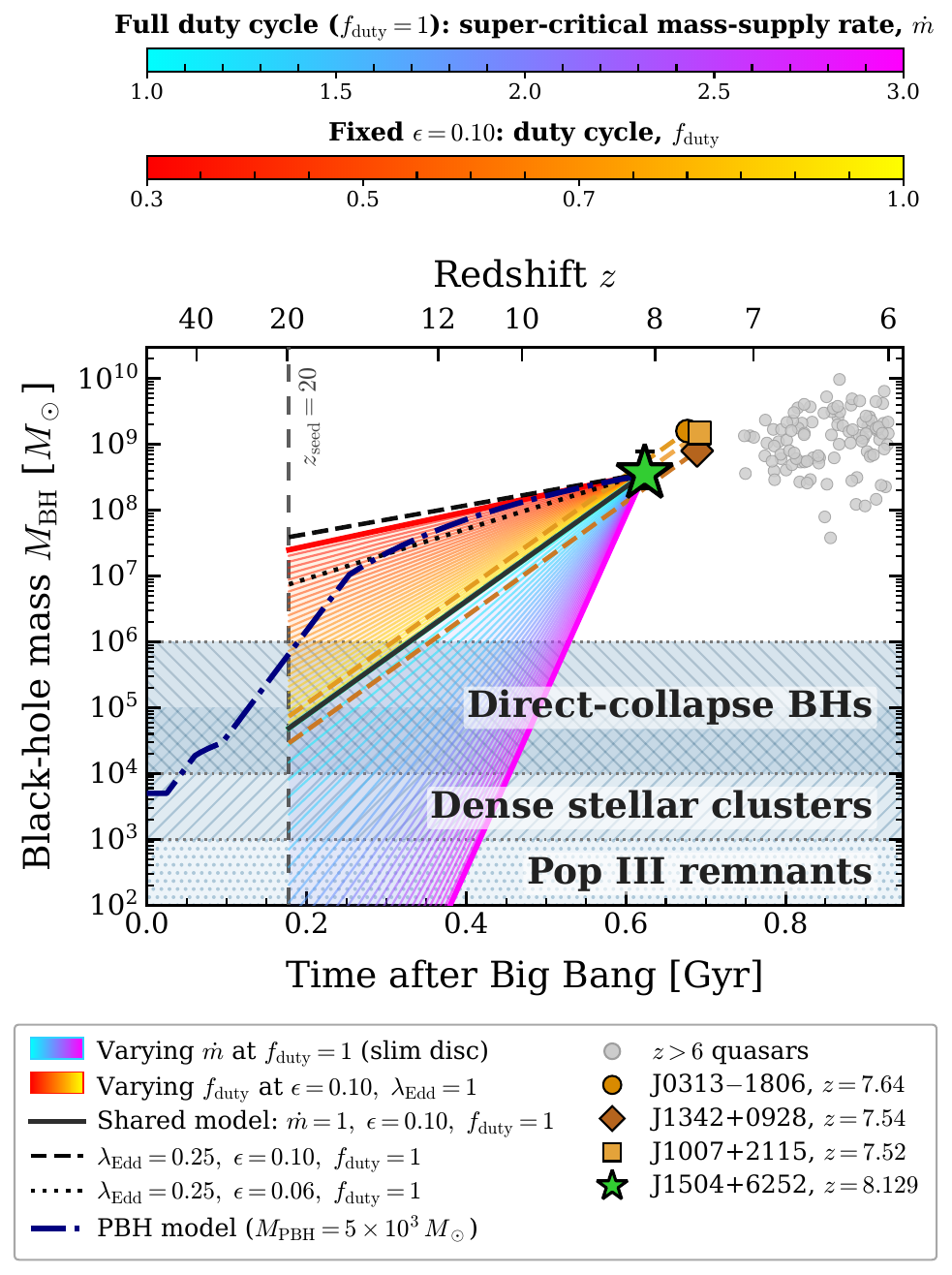}
\caption{Black hole growth curves for J1504+6252 (green star). Quasars at $z>6$ with black hole mass measurements are shown in grey. Orange symbols mark the previously known highest-redshift quasars \citep[$z\simeq7.5$;][]{banados2018a,Yang2020,wang2021}, and the corresponding dashed curves show their growth histories assuming $\epsilon=0.10$, $\lambda_{\rm Edd}=1$, and $f_{\rm duty}=1$ backtraced to $z_\mathrm{seed}=20$. The coloured regions show two families of growth models anchored to the black hole mass estimate for J1504+6252 based on the calibration of \citet{Vestergaard2006}. The error bar spans the central mass estimates obtained from all adopted \hbetashort\ and \mgiishort\ calibrations. The blue--violet models assume continuous super-critical accretion, with $f_{\rm duty}=1$ and $\dot{m}=1.0$--$3.0$, adopting $\lambda_{\rm Edd}\simeq1+\ln\dot{m}$ and the corresponding effective radiative efficiency. The red--yellow models instead assume radiatively efficient, Eddington-limited accretion with $\epsilon=0.10$ and vary the duty cycle over $f_{\rm duty}=0.3$--$1.0$. The solid black curve marks the common limiting model, corresponding to $\dot{m}=1$, $\lambda_{\rm Edd}=1$,  $\epsilon=0.10$, and $f_{\rm duty}=1$. The dashed and dotted black curves show examples of sub Eddington growth with $\lambda_{\rm Edd}=0.25$. The navy dot--dashed curve shows an illustrative primordial-black-hole-seeded assembly history, including halo growth and gas-supply-limited accretion, based on the model of \citet{dayal2024}. Horizontal shaded regions indicate characteristic seed-mass ranges associated with Population~III remnants, runaway stellar collisions in dense clusters, and direct-collapse black holes \citep{inayoshi2020}.}
    \label{fig:models}
\end{figure}

Figure~\ref{fig:models} illustrates the new constraints imposed by the discovery of J1504+6252 using backtraced growth
histories anchored to the measured mass of the quasar. As a physically motivated starting point for the growth histories, we adopt $z_{\rm seed}=20$, representative of the earliest epochs commonly invoked for the formation of massive seeds. PBHs constitute an exception, as they may form much earlier in the early Universe.  For radiatively
efficient accretion, the growth curves are computed using the standard
exponential black hole growth equation
\citep[][see also, e.g. \citealt{volonteri2010}]{salpeter1964},
\begin{equation}
\frac{\diff M_{\rm BH}}{\diff t}
=
\frac{1-\epsilon}{\epsilon}
\frac{\lambda_{\rm Edd}}{t_{\rm Edd}}
f_{\rm duty}\,
M_{\rm BH}\, ,
\end{equation}
where $\epsilon$ is the radiative efficiency, $\lambda_{\rm Edd}$ is the
Eddington ratio during active accretion, $f_{\rm duty}$ is the active
accretion duty cycle, and $t_{\rm Edd}=0.45\,{\rm Gyr}$. For constant
$\epsilon$, $\lambda_{\rm Edd}$, and $f_{\rm duty}$, the solution is
\begin{equation}
M_{\rm BH}(t)
=
M_{\rm seed}
\exp\left[
\frac{1-\epsilon}{\epsilon}
\frac{\lambda_{\rm Edd}f_{\rm duty}}{t_{\rm Edd}}
\left(t-t_{\rm seed}\right)
\right]\, .
\end{equation}

The curves in Fig.~\ref{fig:models} are obtained from this expression and are forced to pass through the observed black hole mass of J1504+6252 at $z=8.13$, or through the measured black hole masses of the three previously most distant quasars at their observed redshifts \citep{banados2018a,Yang2020,wang2021}. 
We explore two families of illustrative black hole growth models. The first assumes radiatively efficient, Eddington-limited accretion, with $\lambda_{\rm Edd}=1$ and a canonical radiative efficiency of $\epsilon=0.10$. Our fiducial model further assumes continuous accretion, $f_{\rm duty}=1$, and is shown for J1504+6252 and the comparison quasars. We then vary the duty cycle over $f_{\rm duty}=0.3$--$1.0$, while keeping $\lambda_{\rm Edd}=1$ and $\epsilon=0.10$, to illustrate the effect of intermittent Eddington-limited growth.

The second family models super-critical accretion, for which the mass
inflow rate exceeds the rate associated with the Eddington luminosity. We
define
\begin{equation}
\dot{m}\equiv\frac{\dot{M}_{\rm inflow}}{\dot{M}_{\rm Edd}}\, ,
\qquad
\dot{M}_{\rm Edd}\equiv\frac{L_{\rm Edd}}{\epsilon c^2}\, ,
\end{equation}
with $\epsilon=0.10$ \citep[e.g.][]{King2016}. At high $\dot{m}$ ($\dot{m}\gg1$), photon trapping, advection, and
winds cause the converted luminosity to increase less rapidly than the mass
inflow rate
\citep{shakura1973,abramowicz1988,poutanen2007,madau2014}. Following the
classical super-critical-disc approximation of \citet[][see also, e.g. \citealt{King2016}]{shakura1973} , we adopt
\begin{equation}
\lambda_{\rm Edd}\simeq1+\ln\dot{m}\, .
\end{equation}
The corresponding effective radiative efficiency is $\epsilon_{\rm eff}=\epsilon\lambda_{\rm Edd}/ \dot{m}$, decreasing from $0.10$ to $0.070$ over the adopted range $\dot{m}=1$--$3$. Much lower efficiencies (as low as $\sim 10^{-3}$)  have been integrated over the growth histories of $z>7$ quasars \citep{davies2019}. We show continuous-growth tracks with $f_{\rm duty}=1$ and $\dot{m}=1$--$3$. The $\dot{m}=1$ case recovers the fiducial Eddington-limited model with $\lambda_{\rm Edd}=1$ and $\epsilon_{\rm eff}=0.10$, while larger $\dot{m}$ produces faster black hole growth with only a modest increase in emergent luminosity.

Within the radiatively efficient models at fixed $\epsilon$, the inferred seed mass is highly sensitive to the duty cycle. Adopting $z_{\rm seed}=20$, continuous Eddington-limited accretion with $\epsilon=0.10$ requires a seed mass of $\sim10^5\,M_\odot$, placing it towards the massive-seed regime and well above the characteristic masses expected for Pop~III remnants. For this model, J1504+6252 provides constraints consistent with those inferred for previously discovered $z\sim 7.5$ quasars \citep{banados2018a,Yang2020,wang2021}.  Progressively lower duty cycles require increasingly massive seeds, moving beyond the characteristic direct-collapse black-hole seed masses considered here. Within the strict assumptions of these toy models and adopting $z_{\rm seed}=20$, duty cycles below approximately $f_{\rm duty}\simeq0.5$ would require seed masses above the populations considered here in order to reproduce the observed mass of J1504+6252 by $z=8.13$.

This constraint can be substantially relaxed by enabling super-critical accretion. Because photon trapping reduces the effective radiative efficiency \citep[e.g.][]{shimura03}, a larger fraction of the supplied rest mass can fall into the black hole without a proportional increase in the emergent luminosity. The super-critical tracks therefore permit substantially lower seed masses, extending into the characteristic Population~III remnant regime. Rapid mass assembly does not necessarily need to occur during archetypical Type 1 UV-luminous quasar phases. The comparison between the black hole mass density at $z\simeq6$ and that inferred from the UV-luminous quasar population indicates that UV-luminous quasars may account for less than ten percent of the early black hole mass growth \citep{jahnke2025a}. This may indicate that most of the growth occurs in obscured systems, radiatively inefficient phases, and potentially
super-critical accretion episodes.  An independent indication for buried growth comes from the discovery of a blazar at $z\sim7$: the implied parent population of randomly aligned jets suggests that a substantial population of rapidly growing, potentially obscured SMBHs may remain undetected in UV-based quasar selections \citep{Banados25}.

Nevertheless, feedback from black hole accretion and supernovae can deplete or expel gas from the shallow gravitational potentials of low-mass seeds, thereby strongly suppressing mass assembly \citep[e.g.][]{johnson2007,whalen2008,milosavljevic2009,alvarez2009,inayoshi2020}. Moreover, continued thin-disc accretion is expected to increase the spin of the black hole and thereby increase its radiative efficiency \citep{thorne1974,volonteri2005b}.  The super-critical tracks should therefore only be interpreted as illustrating
the growth rates required for light seeds to remain viable.

We also show a realisation of PBH-seeded growth based on the assembly model of \citet{dayal2024}. The adopted realisation starts from a $5\times 10^3\,M_\odot$ primordial black hole at matter--radiation equality, which acts as a seed, assembling a dark-matter halo around itself. The black hole remains effectively dormant until the halo becomes sufficiently massive to retain baryons, corresponding to a density of 200 times the critical density. For the black hole seed mass considered here, this corresponds to a redshift of $z \sim 75$. After this, gas can cool to form stars, and the black hole starts growing through gas accretion. Given its low mass, the black hole initially accretes close to the Eddington rate, albeit with fluctuations, down to $z \sim 15$. The Eddington fraction subsequently declines towards lower redshift, from $f_{\rm Edd}\sim1$ at $z\gtrsim20$ to $\sim0.15$ by $z\simeq8$, as growth becomes increasingly supply-limited due to a combination of black hole and supernova feedback. The pronounced change in slope of the track reflects the evolving accretion rate. This scenario likewise provides a viable pathway for assembling a near-billion-solar-mass black hole within the first $630$ million years of cosmic history.

J1504+6252 provides a new benchmark for models of early black hole growth by pushing the required assembly to still earlier times. However, the discovery of J1504+6252 does not enable a robust distinction between these formation pathways.

Strikingly, J1504+6252 has a comparatively low Eddington ratio relative to other known $z>6$ quasars (Fig.~\ref{fig:mbhlbol}). From the
\hbetashort-based black hole mass and bolometric luminosity, we infer 
$\lambda_{\rm Edd}=0.26^{+0.05}_{-0.04}$ from the calibration of \citet{Vestergaard2006} and a range $\lambda_{\rm Edd}=0.12\text{--}0.44$ from all single-epoch calibrations. Thus, although the SMBH is already massive, it is not currently accreting close to the Eddington limit. 
The discovery of a sub-Eddington quasar with $\logten(L_\mathrm{bol}/[\mathrm{erg\,s^{-1}}])\sim46$ at $z>8$ therefore suggests that substantially brighter members of the same black hole population may remain to be discovered if they are not enshrouded in dust. At fixed black hole mass, increasing the accretion rate from the benchmark value $\lambda_{\rm Edd}\simeq0.26$ to $\lambda_{\rm Edd}=1$ would increase the luminosity by a factor of $3.8$, corresponding to a brightening of $1.46$ mag. An Eddington-limited analogue of J1504+6252 would thus have approximately $\JE=21.5$,  $\HE=21.2$, and $M_{1450}=-25.5$, making it easily accessible to \Euclid.
If J1504+6252 is representative of the emerging $z>8$ population, its low $\lambda_{\rm Edd}$ would imply that some billion-solar-mass black holes had already completed their most rapid growth phase by this epoch. In either case, the presence of such a
massive black hole in a comparatively low-accretion state requires efficient mass assembly at even earlier times, followed by a decline to the observed sub-Eddington phase.

\section{Conclusions}
\label{sec:conclusions}
We have presented the discovery of the quasar J1504+6252, a source we name Dún Aonghasa. Initially selected from \textit{Euclid} imaging, J1504+6252 required a sequence of ground- and space-based NIR spectroscopic observations to confirm its nature as the most distant quasar known to date. From the narrow \hbetashort emission line, we measure a systemic redshift of $z=8.129\pm0.005$, pushing the quasar redshift frontier about 39\,Myr (or by 6\%) earlier in cosmic history than the previous record holder at $z=7.77$ \citep{yang2026}. Based on the width of the broad \hbetashort\ emission line, we infer a benchmark single-epoch black hole mass of $\logten(M_{\rm BH}/M_\odot)=8.56^{+0.43}_{-0.43}$ using the \citet{Vestergaard2006} calibration, while the central values obtained from all adopted \hbetashort\ and \mgiishort\ calibrations span $\logten(M_{\rm BH}/M_\odot)=8.33$--$8.89$. The corresponding Eddington ratios span $\lambda_{\rm Edd}=0.12$--$0.44$, indicating that the quasar is accreting at a comparatively low rate relative to much of the known $z>6$ quasar population. Dún Aonghasa establishes a new quasar redshift frontier as the first confirmed quasar at $z>8$, providing a powerful new benchmark for early SMBH growth.

The comparatively low instantaneous accretion rate of J1504+6252 extends the emerging picture of the earliest quasar population. Its large black hole mass shows that substantial growth must have occurred before the epoch at which the source is observed, whereas its current bolometric luminosity is not consistent with a phase of rapid mass assembly. The discovery of J1504+6252 implies that early massive black holes may occupy a broader range of accretion states than is represented by the predominantly luminous, rapidly accreting systems identified at $z>6$ to date.

\citet{Barnett-EP5} predicted that \textit{Euclid} should identify approximately 8--23 quasars at $z>8$ over a survey area of $\sim$14\,000 deg$^2$, with the range driven primarily by assumptions about the evolution of the quasar luminosity function at the highest redshifts. Scaling this prediction to the $\sim$3644 deg$^2$ considered here yields an expectation of $\sim$2--6 quasars. However, the extensive spectroscopic follow-up campaign required to discover J1504+6252 highlights the practical challenges of identifying such objects should there indeed be more to discover. As the survey area and follow-up coverage expand, \Euclid will provide a clearer picture of the size of the $z>8$ quasar population, and thereby provide unparalleled constraints on the abundance and growth of the earliest massive black holes.

\begin{acknowledgements}
We thank the anonymous referee for their constructive comments and suggestions, which have helped us improve the clarity and presentation of the paper.
We thank Joseph Shields, Dave Thompson, and the staff of the Large Binocular Telescope (LBT) for their support of this work.
The authors also gratefully acknowledge the assistance of our primary STScI support contacts, Patricia Royle and Alison Vick, as well as Amanda Pagul, Anton Koekemoer, and Cheryl Pavlovsky, whose support made the discovery observations possible.
This research was supported by the Excellence Cluster ORIGINS, which is funded by the Deutsche Forschungsgemeinschaft (DFG, German Research Foundation) under Germany's Excellence Strategy -- EXC 2094/2 -- 390783311. 
RD acknowledges support from the INAF RF 2024 mini-grant  ``The interstellar medium at high redshift'' and from the PRORIS 2025 ``COSMOWEBB''. 
MC acknowledges support from INAF RF 2024 GO grant ``Revealing the nature of bright galaxies at cosmic dawn with deep JWST spectroscopy''.
\AckEC
Some of the data presented herein were obtained at Keck Observatory, which is a private 501(c)3 non-profit organisation operated as a scientific partnership among the California Institute of Technology, the University of California, and the National Aeronautics and Space Administration. The Observatory was made possible by the generous financial support of the W. M. Keck Foundation.
The authors wish to recognise and acknowledge the very significant cultural role and reverence that the summit of Maunakea has always had within the Native Hawaiian community. We are most fortunate to have the opportunity to conduct observations from this mountain.
This research is based on observations made with the NASA/ESA \textit{Hubble} Space Telescope obtained from the Space Telescope Science Institute, which is operated by the Association of Universities for Research in Astronomy, Inc., under NASA contract NAS 5–26555. These observations are associated with programme18112 (PI: J.~Wolf).
This paper includes data from the LBT (programmeID: MPIA-2026A-001, PI E.~Ba\~{n}ados).
The LBT is an international collaboration among institutions in the United States and Europe. At the time data were acquired for this research, LBT Corporation Members were the University of Arizona on behalf of the Arizona Board of Regents; Istituto Nazionale di Astrofisica, Italy; and The Ohio State University, representing The Ohio State University, University of Notre Dame, University of Minnesota, and University of Virginia.  This research used the facilities of the Italian Center for Astronomical Archives (IA2) operated by INAF at the Astronomical Observatory of Trieste.  Observations have benefited from the use of ALTA Center (alta.arcetri.inaf.it) forecasts performed with the Astro-Meso-Nh model. Initialization data of the ALTA automatic forecast system come from the General Circulation Model (HRES) of the European Centre for Medium Range Weather Forecasts.
This work is based in part on observations made with the NASA/ESA/CSA \textit{James Webb} Space Telescope. The data were obtained from the \textit{Mikulski} Archive for Space Telescopes at the Space Telescope Science Institute, which is operated by the Association of Universities for Research in Astronomy, Inc., under NASA contract NAS 5-03127 for JWST. These observations are associated with program \#12761 (PI: J.~Wolf).
This research made use of {\ttfamily PypeIt}\footnote{\url{https://pypeit.readthedocs.io/en/stable/}} version
1.17.3, a Python
package for semi-automated reduction of astronomical slit-based spectroscopy. 
Funding for the Sloan Digital Sky Survey V has been provided by the Alfred P. Sloan Foundation, the Heising-Simons Foundation, the National Science Foundation, and the Participating Institutions. SDSS acknowledges support and resources from the Center for High-Performance Computing at the University of Utah. SDSS telescopes are located at Apache Point Observatory, funded by the Astrophysical Research Consortium and operated by New Mexico State University, and at Las Campanas Observatory, operated by the Carnegie Institution for Science. The SDSS web site is \url{www.sdss.org}.
SDSS is managed by the Astrophysical Research Consortium for the Participating Institutions of the SDSS Collaboration, including the Carnegie Institution for Science, Chilean National Time Allocation Committee (CNTAC) ratified researchers, Caltech, the Gotham Participation Group, Harvard University, Heidelberg University, The Flatiron Institute, The Johns Hopkins University, L'Ecole polytechnique f\'{e}d\'{e}rale de Lausanne (EPFL), Leibniz-Institut f\"{u}r Astrophysik Potsdam (AIP), Max-Planck-Institut f\"{u}r Astronomie (MPIA Heidelberg), Max-Planck-Institut f\"{u}r Extraterrestrische Physik (MPE), Nanjing University, National Astronomical Observatories of China (NAOC), New Mexico State University, The Ohio State University, Pennsylvania State University, Smithsonian Astrophysical Observatory, Space Telescope Science Institute (STScI), the Stellar Astrophysics Participation Group, Universidad Nacional Aut\'{o}noma de M\'{e}xico, University of Arizona, University of Colorado Boulder, University of Illinois at Urbana-Champaign, University of Toronto, University of Utah, University of Virginia, Yale University, and Yunnan University.  
This work made use of Astropy\footnote{\url{https://www.astropy.org}}, a community-developed core Python package and an ecosystem of tools and resources for astronomy.

\end{acknowledgements}

\bibliographystyle{aa}
\bibliography{merged_bibliography}

\begin{appendix}

\onecolumn
\section{Lower redshift SDSS quasar analogues}
\label{sec:cousins}

\begin{figure*}[h!]
    \centering
    \includegraphics[]{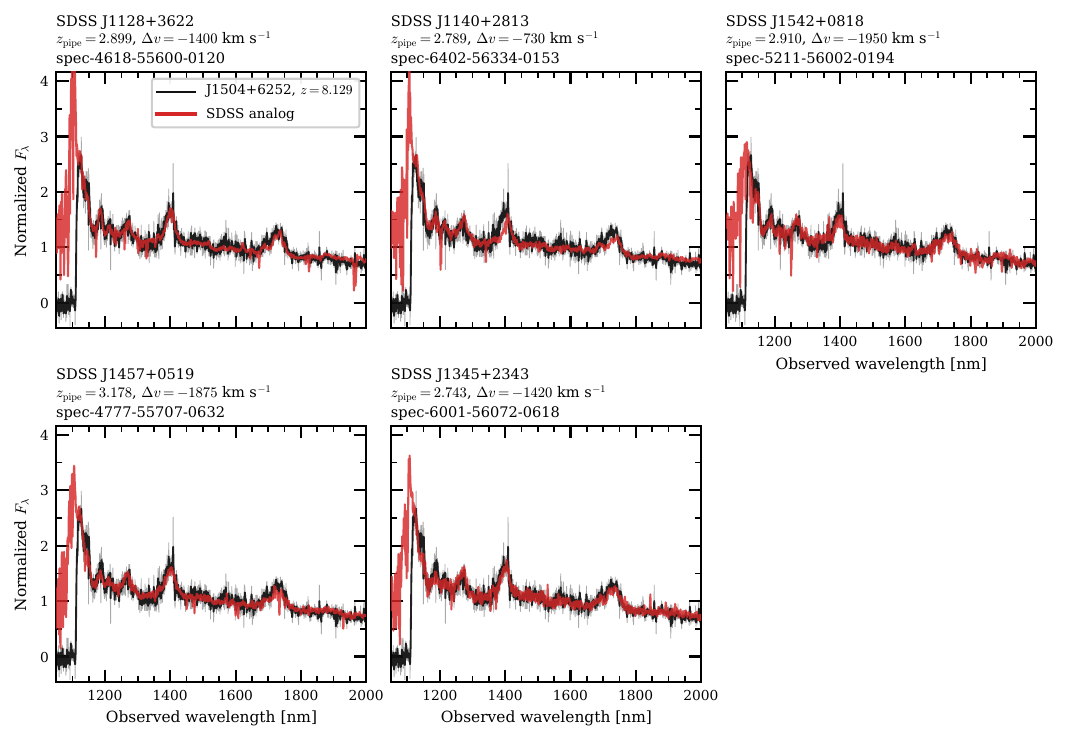}
    \caption{Comparison of the rest-frame UV spectrum of J1504+6252 (black) with five lower-redshift SDSS quasars (red) selected as its nearest spectral neighbours. Candidate analogues were identified by minimising the Euclidean distance between continuum-normalised spectra over the common rest-frame wavelength interval 1230--2000\,\AA. For each comparison quasar, a small velocity offset was additionally fitted to account for intrinsic broad-line centroid shifts before scaling the continuum. The displayed spectra are normalised over the wavelength range 1700--1800\,\AA. Despite the difference in cosmic age, the overall UV spectral properties of J1504+6252 are remarkably similar to those of several quasars at lower redshift, including the broad emission-line profiles and continuum shape.}
    \label{fig:cousins}
\end{figure*}

We search for the most similar quasars to J1504+6252 within the data release 14 Quasar catalogue (DR14Q) from the extended Baryon Oscillation Spectroscopic Survey (eBOSS) of the Sloan Digital Sky Survey IV (SDSS-IV, \citealt{Paris18}), which we access via \texttt{igmspec} \citep{Prochaska17}. The nearest-neighbour search and the parent sample are the same as described in \citet{Bosman21}. In order to cover the spectrum from the \lya up to the \ciii emission line, we select quasars with $2.7<z<3.5$ with spectral ${\rm S/N} > 7$ and no presence of broad absorption lines; this yields 9194 low-$z$ quasars. The continua are compared over $123$\,nm $<\lambda_\text{rest}<200$\,nm \ in order to avoid the Lyman-$\alpha$ absorption damping wing in J1504+6252. The spectra are automatically fit with a spline following the procedure outlined by \citet{Young79} and \citet{Carswell1982}, first implemented by \citet{DallAglio08} and with the improvements of \citet{Davies2018b} and \citet{Bosman21}. Starting from an initial estimated spline spacing which is equal in velocity space for both J1504+6252 and the low-$z$ sample, the procedure iteratively rejects poor pixels, masks regions affected by absorption, and optimises the placement of spline nodes. Finally, the distances between optimal spline fits were computed using the \texttt{NearestNeighbours} module of \texttt{sklearn} \citep{sklearn} in Euclidean distance. The top $5$ nearest neighbours of J1504+6252 within eBOSS have Euclidean distances $d<3$, which is also typical of the low-$z$ quasar population. The identified analogues are SDSS J1128+3622 (plate--MJD--fibre 4618--55600--0120), SDSS J1140+2813 (6402--56334--0153), SDSS J1542+0818 (5211--56002--0194), SDSS J1457+0519 (4777--55707--0632), and SDSS J1345+2343 (6001--56072--0618).
These five best nearest neighbours are shown in Fig.~\ref{fig:cousins}. We note that since the nearest neighbour search stops at $\lambda_\text{rest}=2000\,$\AA, the spectrum is not shown past this value.  In the spectral-property catalogue of \citet{wushen2022}, these quasars span redshifts of $z=2.75$--$3.18$ and single-epoch \civshort-based black-hole masses of $\log_{10}(M_{\rm BH}/M_\odot)=9.41$--$9.82$. Applying the bolometric corrections adopted by \citet{wushen2022} to the rest-frame UV continuum luminosities yields bolometric luminosities of $\logten[L_{\rm bol}/{\rm (erg\,s^{-1})}]=46.71$--$47.53$, corresponding to a range of Eddington rates $\lambda_\mathrm{Edd}=0.12-0.47$. These analogues are thus more luminous than J1504+6252, but their nominal Eddington rates are comparable.

\onecolumn
\section{Line measurements}
\label{sec:appendix}
In Table~\ref{tab:ap}, we detail the decomposition of the emission lines used for the estimation of black hole masses in this work. 

\begin{table*}[h!]
\caption{Emission-line measurements for J1504+6252.
Uncertainties correspond to the 16th and 84th percentiles of the
posterior distributions. The [\ion{O}{iii}] values are one-sided
posterior $3\sigma$ upper limits.}
\label{tab:j1504_line_measurements}
\centering
\footnotesize
\setlength{\tabcolsep}{3.2pt}
\renewcommand{\arraystretch}{1.10}

\begin{tabularx}{\textwidth}{
    @{}
    l
    l
    c
    c
    c
    c
    >{\raggedright\arraybackslash}X
    @{}
}
\toprule
\toprule

Feature &
Component &
FWHM &
$\Delta v$ &
Flux &
$\mathrm{EW}_{0}$ &
Comment \\
&
&
[$\mathrm{km\,s^{-1}}$] &
[$\mathrm{km\,s^{-1}}$] &
[$10^{-17}\,\mathrm{erg\,s^{-1}\,cm^{-2}}$] &
[\si{\angstrom}] &
\\
\midrule

\multicolumn{7}{l}{\hbetashort--\oiiishort\ complex} \\

\hbetashort &
Narrow &
$427^{+153}_{-114}$ &
$0$ &
$0.316^{+0.080}_{-0.071}$ &
$5.90^{+1.51}_{-1.30}$ &
\\

\hbetashort &
Intermediate &
$2294^{+140}_{-216}$ &
$-261^{+90}_{-90}$ &
$1.344^{+0.171}_{-0.176}$ &
$25.12^{+3.30}_{-3.36}$ &
\\

\hbetashort &
Broad base &
$7801^{+532}_{-556}$ &
$+744^{+110}_{-179}$ &
$2.750^{+0.216}_{-0.212}$ &
$51.41^{+4.29}_{-4.13}$ &
\\

\hbetashort &
BLR total &
$3159^{+269}_{-272}$ &
$+397^{+92}_{-92}$ &
$4.096^{+0.192}_{-0.212}$ &
$76.58^{+4.13}_{-4.68}$ &
Broad and intermediate components \\

\hbetashort &
All components &
$\cdots$ &
$+370^{+82}_{-87}$ &
$4.413^{+0.189}_{-0.189}$ &
$82.44^{+4.30}_{-4.17}$ &
BLR and narrow components \\

Optical \feii &
4434--4684\,\si{\angstrom} &
$1267^{+194}_{-146}$ &
$0$ (fixed) &
$3.134^{+0.212}_{-0.213}$ &
$51.55^{+4.01}_{-4.08}$ &
Additional convolution;
$R_{\rm FeII}=0.766^{+0.055}_{-0.050}$ \\

[\ion{O}{iii}] $\lambda4959$ &
Narrow &
$427^{+153}_{-114}$ &
$0$ (tied) &
$<0.0237$ &
$<0.463$ &
$3\sigma$ upper limit \\

[\ion{O}{iii}] $\lambda5007$ &
Narrow &
$427^{+153}_{-114}$ &
$0$ (tied) &
$<0.0706$ &
$<1.407$ &
$3\sigma$ upper limit \\

[\ion{O}{iii}] doublet &
Total &
$427^{+153}_{-114}$ &
$0$ (tied) &
$<0.0943$ &
$<1.870$ &
$3\sigma$ upper limit \\

\midrule
\multicolumn{7}{l}{\mgiishort\ region} \\

\mgiishort &
Broad component &
$3623^{+162}_{-173}$ &
$-733^{+58}_{-57}$ &
$2.158^{+0.081}_{-0.083}$ &
$14.72^{+0.68}_{-0.66}$ &
Single-component fit \\

\mgiishort &
BLR doublet profile &
$3739^{+157}_{-170}$ &
$-733^{+58}_{-57}$ &
$2.158^{+0.081}_{-0.083}$ &
$14.72^{+0.68}_{-0.66}$ &
FWHM used for $M_{\rm BH}$ \\

UV \feii &
2520--3178\,\si{\angstrom} &
$1550^{+265}_{-198}$ &
$0$ (fixed) &
$13.690^{+1.049}_{-0.845}$ &
$93.39^{+8.53}_{-6.60}$ &
Final effective FWHM; integrated over the fitted window \\

\bottomrule
\end{tabularx}

\tablefoot{
The emission-line FWHM values are intrinsic and account for the
instrumental line-spread function. Velocity shifts are measured relative
to the adopted systemic reference, with positive values indicating
redshifts. The total \hbetashort\ BLR profile is the sum of the
intermediate and broad-base components, excluding the narrow core.
The optical \feii FWHM is the additional Gaussian convolution applied
to the native empirical template and should not be interpreted as the
intrinsic \feii line width. The UV \feii FWHM is the final
effective template width, accounting for the adopted native template
FWHM of $900\,\mathrm{km\,s^{-1}}$. The [\ion{O}{iii}] widths and
velocity shifts are tied to the narrow \hbetashort\ component, with
$F_{5007}/F_{4959}=2.98$. The quoted line-flux and equivalent-width
uncertainties are derived from the spectral-fit posterior and do not
include the 5\% absolute-flux-calibration systematic, which is included
in the luminosity measurements.
}
\label{tab:ap}

\end{table*}

\end{appendix}

\end{document}